\documentclass[fleqn,usenatbib]{mnras}

\usepackage{newtxtext,newtxmath}
 
\usepackage[T1]{fontenc}

\DeclareRobustCommand{\VAN}[3]{#2}
\let\VANthebibliography\thebibliography
\def\thebibliography{\DeclareRobustCommand{\VAN}[3]{##3}\VANthebibliography}

\usepackage{graphicx}	% Including figure files
\usepackage{amsmath}	% Advanced maths commands
\usepackage{amssymb}	% Extra maths symbols
\usepackage{multirow} % Required for multirows
\usepackage{ctable} % for \specialrule command
\usepackage{textcomp}
\usepackage{placeins}
\usepackage{float}
\usepackage{adjustbox}

\newcommand{\FB}[1]{\textcolor{orange}{}}

\newcommand{\aperp}{\alpha_\perp}
\newcommand{\apar}{\alpha_\parallel}
\newcommand{\ihMpc}{$h^{-1}$Mpc}

\usepackage{caption}

\usepackage{comment}
\usepackage{tablefootnote}
\usepackage{threeparttable}
\usepackage[normalem]{ulem}  % for strikethrough
\title[Observational systematics on the BAO]{
The clustering of the SDSS-IV extended Baryon Oscillation Spectroscopic Survey quasar sample: Testing observational systematics on the Baryon Acoustic Oscillation measurement}

\author[G. Merz et al.]
{Grant Merz,$^{1}$\thanks{E-mail: gm240915@ohio.edu}
Mehdi Rezaie,$^{1}$
Hee-Jong Seo,$^{1}$
Richard Neveux$^{2}$, 
Ashley J. Ross$^{3}$,
\newauthor Florian Beutler$^{4}$, Will J. Percival$^{5,6,7}$, Eva Mueller$^{8}$, H\'ector Gil-Mar\'in$^{9,10}$, 
\newauthor
Graziano Rossi$^{11}$,
Kyle Dawson$^{12}$,
Joel R. Brownstein$^{12}$
Adam D. Myers$^{13}$,
\newauthor
Donald P. Schneider$^{14,15}$,
Chia-Hsun Chuang$^{16}$,
Cheng Zhao$^{17}$,
Axel de la Macorra$^{18}$,
\newauthor
Christian Nitschelm$^{19}$
\\~\\
$^{1}$Department of Physics and Astronomy, Ohio University, Athens, OH 45701, USA\\
$^{2}$IRFU,CEA, Universit\'e Paris-Saclay, F-91191 Gif-sur-Yvette, France\\
$^{3}$Center of Cosmology and AstroParticle Physics, The Ohio State University, Columbus, OH 43210, USA\\
$^{4}$School of Physics and Astronomy, University of Edinburgh, Edinburgh, EH9 3FD, United Kingdom\\
$^{5}$Waterloo Centre for Astrophysics, University of Waterloo, Waterloo, ON N2L 3G1, Canada\\
$^{6}$Department of Physics and Astronomy, University of Waterloo, Waterloo, ON N2L 3G1, Canada\\
$^{7}$Perimeter Institute for Theoretical Physics, 31 Caroline St. North, Waterloo, ON N2L 2Y5, Canada\\
$^{8}$Department of Physics, University of Oxford, Denys Wilkinson Building, Keble Road, Oxford OX1 3RH, UK\\
$^{9}$Institut de Ci\`encies del Cosmos, Universitat de Barcelona, ICCUB, Mart\'i i Franqu\`es 1, E08028 Barcelona, Spain\\ 
$^{10}$Institut  d'Estudis  Espacials  de  Catalunya  (IEEC),  E08034  Barcelona,  Spain\\ 
$^{11}$Department of Physics and Astronomy, Sejong University, Seoul 143-747, Korea\\
$^{12}$Department of Physics and Astronomy, University of Utah, 115 S. 1400 E., Salt Lake City, UT 84112, USA\\
$^{13}$Department of Physics and Astronomy, University of Wyoming, Laramie, WY. 82071\\
$^{14}$Department of Astronomy and Astrophysics, The Pennsylvania State University, University Park, PA 16802, USA\\
$^{15}$Institute for Gravitation and the Cosmos, The Pennsylvania State University, University Park, PA 16802, USA \\ 
$^{16}$Kavli Institute for Particle Astrophysics and Cosmology, Stanford University, 452 Lomita Mall, Stanford, CA 94305, USA\\
$^{17}$Institute of Physics, Laboratory of Astrophysics, \'Ecole Polytechnique F\'ed\'erale de Lausanne (EPFL), Observatoire de Sauverny, CH-1290 Versoix, Switzerland\\
$^{18}$Instituto de Física, Universidad Nacional Autónoma de México, Apdo. Postal 20-364, México\\
$^{19}$ Centro de Astronom\'a, Universidad de Antofagasta, Avenida Angamos 601, Antofagasta 1270300, Chile\\
}

\date{Accepted XXX. Received YYY; in original form ZZZ}

\pubyear{2020}

\DeclareUnicodeCharacter{2212}{-}
\DeclareUnicodeCharacter{2032}{'}

\begin{document}
\label{firstpage}
\pagerange{\pageref{firstpage}--\pageref{lastpage}}
\maketitle

\begin{abstract}

Baryon Acoustic Oscillations are considered to be a very robust standard ruler against various systematics. This premise has been tested against observational systematics, but not to the level required for the next generation of galaxy surveys such as the Dark Energy Spectroscopic Instrument (DESI) and Euclid. In this paper, we investigate the effect of observational systematics on the BAO measurement of the final sample of quasars from the extended Baryon Oscillation Spectroscopic Survey Data Release 16 in order to prepare and hone a similar analysis for upcoming surveys. We employ catalogues with various treatments of imaging systematic effects using linear and neural network-based nonlinear approaches and consider how the BAO measurement changes. We also test how the variations to the BAO fitting model respond to the observational systematics. As expected, we confirm that the BAO measurements obtained from the DR16 quasar sample are robust against imaging systematics well within the statistical error, while reporting slightly modified constraints that shift the line-of-sight BAO signal by less than 1.1\% . We use realistic simulations with similar redshift and angular distributions as the DR16 sample to conduct statistical tests for validating the pipeline, quantifying the significance of differences, and estimating the expected bias on the BAO scale in future high-precision data sets.   Although we find a marginal impact for the eBOSS QSO data, the work presented here is of vital importance for constraining the nature of dark energy with the BAO feature in the new era of big data cosmology with DESI and Euclid.
\end{abstract}


%%%%%%%%%%%%%%%%%%%%%%%%%%%%%%%%%%%%%%%%%%%%%%%%%%

%%%%%%%%%%%%%%%%% BODY OF PAPER %%%%%%%%%%%%%%%%%%

\section{Introduction}
Today, dark energy, a mysterious component behind the accelerating cosmic expansion is one of the leading focuses of cosmology research \citep{Weinberg13}. The Baryon Acoustic Oscillations (BAO) feature in the large-scale clustering of luminous matter measured by galaxy surveys \citep[e.g.,][]{Eisenstein05,Ata17} is one of the most robust probes of the expansion rate/history of the Universe and thus the nature of dark energy. 
To date, the BAO signal has been studied by measuring the spatial clustering of several different tracers of dark matter in various phases of the Sloan Digital Sky Survey \citep[SDSS,][]{york2000sloan}, e.g. SDSS-II, the Baryon Oscillation Spectroscopic Survey \citep[BOSS;][]{BOSS13}, and extended BOSS \citep[eBOSS;][]{eBOSS16}, a part of SDSS-IV \citep{SDSS4}. Measurements have reached out to redshifts of $z=2.2$.  The most accurate BAO measurements of any redshift survey thus far are presented in \citet{eBOSS2020}.

The two primary categories of systematic error in the BAO analysis are theoretical and observational systematics. Theoretical systematics are caused by our incomplete understanding of the mechanisms behind structure growth, peculiar velocity and galaxy formation, while observational systematics are primarily associated with varying imaging properties, inaccurate photometric calibrations and redshift measurements.
Observational realities, in particular imaging properties such as Galactic extinction, stellar contamination, seeing, survey depths, etc., are known to introduce spurious density fluctuations in the galaxy and quasar or quasi-stellar object (QSO) sample on large scales in the spectroscopic target selection stage \citep[e.g.,][]{Thomas2011PhRvL.106x1301T}. Such clustering contamination tends to be increasingly severe as we focus on a larger scale and indeed it is a primary limitation in cosmology analyses of primordial non-Gaussianity~\citep[e.g.,][]{Pullen13, Ross13} or tests of the relativistic effect \citep[e.g.,][]{Wang2020MNRAS.499.2598W}.  The scale-dependent feature in the clustering signal that is inherent to these phenomena is degenerate with the contamination due to observational systematic effects. 
In this respect, mock challenges designed to validate data analysis pipelines and assess the impact of systematics in massive data sets such as the ones developed for the final eBOSS Data Release 16 (DR16) analyses \citep[i.e.,][]{Rossi20,Smith20,Alam20} are crucial assets for large-volume surveys.

The effect of observational systematics is generally mitigated by a regression analysis of the correlation between the observed galaxy density and a set of templates that describe imaging properties due to observation \citep[e.g.,][]{bautista2018sdss}. The end product is often a selection mask that can maximally remove the correlation by appropriately weighting galaxies or by removing the power spectrum modes that are mostly affected by the unwanted feature  \citep[e.g.,][]{kalus2019map}. It is rather ambiguous how to judge how much of the systematics have been removed after this mitigation step without making certain assumptions on the expected cosmological signal. One could estimate the residual based on the remaining dependence of the mitigated density fluctuation on various imaging systematics or by investigating the effect on the 2-point clustering before and after mitigation, i.e., by observing the convergence behaviour.

Despite their detrimental effect on, e.g. primordial non-Gaussianity, for observational systematics to cause a disruption in the BAO signal, they must peak closely at the scale of the BAO.  The finding so far has been that such spurious observational effects become negligible for the typical scale and shape of Baryon Acoustic Oscillations after a reasonable mitigation treatment, or even without any mitigation treatment. \citet{Ross2017} conducted a careful analysis of the observational systematic effect on the BAO measurement of the BOSS DR12 massive galaxies between $0.2<z<0.75$. They identified a dependence of galaxy fluctuation on stellar density and seeing, conducted a linear regression, and constructed a selection function. The effect on the BAO fit before and after the mitigation was less than 10-15\% of the quoted statistical precision of 1.3\% and 2.4\%. \citet{Neveux20} conducted the BAO analysis using the final sample of quasars \citep{lyke2020ApJS..250....8L, Ross2020} from the eBOSS DR16  \citep{Ahumada2020ApJS..249....3A}. For these data, the selection function was derived using a multivariate linear regression of the quasar density against four imaging attributes including Galactic extinction, seeing in i-band, sky brightness in i-band, and survey depth in g-band. The change in the BAO measurement before and after the mitigation is about 30\% of the statistical precision, which is still reasonable when accounting for the statistical scatter due to a stronger observational systematic effects of this sample compared to that of the BOSS data.

Meanwhile, this default treatment still leaves a spurious signal on very large scales ($k<0.01$~\ihMpc) in the power spectrum of the eBOSS DR16 QSO sample, which could imply either a large amount of residual systematic effect or a substantial amount of primordial non-Gaussianity \citep[see, e.g.,][for a discussion]{mueller2021}. Our companion paper, \cite{rezaie2021} developed a neural network (NN)-based method for this sample to account for nonlinear imaging systematics, and tested the density fluctuations against 17 different and highly correlated imaging properties. In the process, we found that the stellar density is one of the primary sources of systematics error, and it is critical to include a stellar density template constructed from the Gaia spacecraft data \citep{gaia2018}. The paper shows that with the default mitigation and Poisson statistics to account for the sparsity of the QSO sample, the residual $\chi^2$ of density fluctuation against 17 imaging attributes in fact was still too high compared to what is expected for purely cosmological fluctuations of the benchmark model and that the NN-based method can substantially reduce the $\chi^2$ into a reasonable range and subsequently reduce the spurious power on very large scales.

\begin{figure}
    \centering
    \includegraphics[width=\columnwidth]{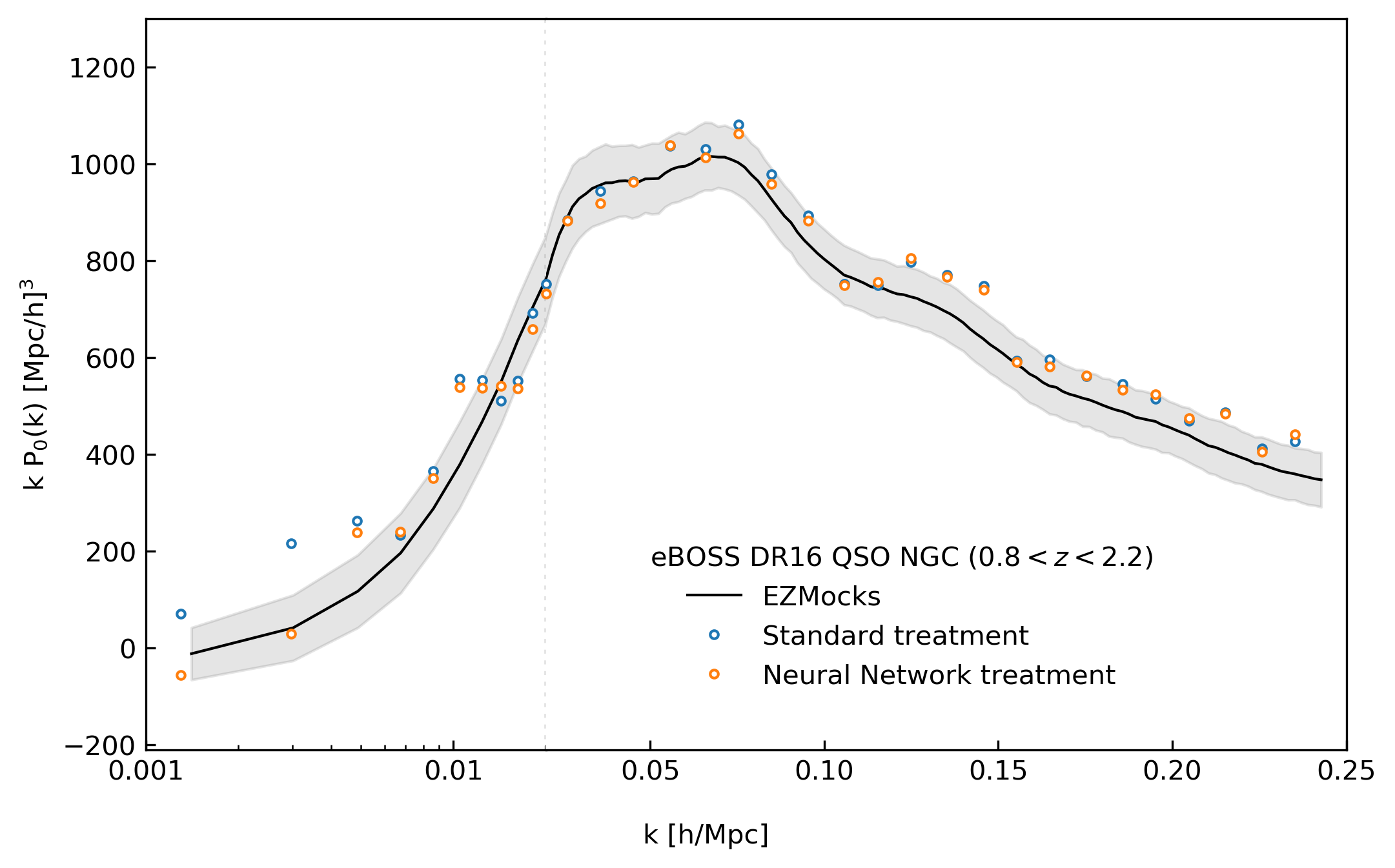}
    \caption{A comparison of the power spectrum of the DR16 QSO sample in the NGC region when treated with default (linear) systematic mitigation vs a NN mitigation (data points).  Also shown is a black line indicating the mean of EZmocks with a 1$\sigma$ band of one mock.}
    \label{fig:powercomp}
\end{figure}

Figure \ref{fig:powercomp} shows a comparison of the NGC power spectrum of the DR16 QSO sample when treated with linear or NN mitigations. While it is qualitatively clear that the excess clustering becomes more prominent at $k < 0.01~h/{\rm Mpc}$ and the new method/weights being necessary for an accurate cosmology measurement at least on very large scales, it is important to test and confirm that these weights do not substantially alter power spectrum over $k >0.01~h/{\rm Mpc}$ and to quantify any impacts on the statistical uncertainty of the BAO. This confirmation will also imply that we can consistently conduct the analysis of the BAO using the catalogue that is optimal for the primordial non-Gaussianity measurement. This particular form of systematic error, i.e., the effect of Gaia stellar density and the nonlinear model has not been tested for the BAO measurement in the form of the convergence test. 

In this paper, we therefore use the new NN-based selection masks from \citep{Rezaie19} for testing the robustness of the BAO measurement from the eBOSS DR16 QSO data in \citet{Neveux20} and for quantifying the effects of the systematics when the scatter in the mitigation stage is propagated to the final constraint. We also investigate additional freedom in the BAO fitting compared to the default used in \citet{Neveux20} and observe the interplay between the observational systematics and the freedom in the BAO fitting model. We will show that the eBOSS QSO BAO measurement is again robust against observational systematics, given the moderate statistical precision.
We will also show that introducing an additional freedom in the BAO fitting can improve the constraint slightly. While the main result of this paper is a confirmation that BAO feature is robust against observational systematics given the statistical precision of the eBOSS QSO sample, the test we lay down here will be increasingly more important and indispensable for upcoming surveys that are reaching wider and deeper (in magnitude), as such surveys are potentially subject to much more significant systematic effect than we study in this paper.

In addition to the systematic test of the main QSO sample, in Appendix \ref{App:highz}, we also present a simple BAO signal extraction from the eBOSS QSO high-z sample in the redshift range $2.2<z<3.5$ that was used for the Lyman-$\alpha$ (Ly-$\alpha$) forest BAO measurement, in order to test the feasibility of adding the QSOs for future Ly-$\alpha$ BAO measurements. While we could not extract the BAO from the eBOSS high-z sample alone, when combined with the main sample, we find a slight improvement, i.e. \char`~10\% of the BAO constraint due to the very high shot noise of this sample. 

This paper is structured as following. In \S~\ref{sec:data}, we summarize the specifics of the eBOSS DR16 QSO measurement. In \S~\ref{sec:methods}, we summarize the method of deriving observed power spectrum, constructing the power spectrum model, the BAO fitting method. Also, we summarize the standard and NN-based systematic mitigation methods. In \S~\ref{sec:results}, we present the effect of the various mitigation strategies on the BAO best fit and errors, the effect of propagating the error associated with the mitigation and the interplay between the mitigation method and the BAO fitting parameters. We also include a mock test to understand the significance of the difference made due to different mitigation strategies. Finally in \S~\ref{sec:conclusion}, we conclude. 

\section{Data}\label{sec:data}
This paper uses the large-scale clustering catalogues for the eBOSS DR16 QSO sample \citep{lyke2020ApJS..250....8L}, which are presented in \cite{Ross2020} and further enhanced in our companion paper \citep{rezaie2021}. The targeting of eBOSS QSOs is described in \cite{myers2015ApJS..221...27M}, and utilized optical and infrared imaging, respectively, from SDSS \citep{york2000sloan, eisenstein2011sdss} and WISE \citep[Wide-Field Infrared Survey Explorer;][]{wright2010wide}. Using the BOSS double-armed spectrogrographs \citep{smee2013multi}, spectroscopic data was taken from  fibers on the plates of the 2.5-meter Sloan Telescope \citep{SloanTelescope} at Apache Point Observatory in New Mexico. The DR16 sample takes advantage of Legacy objects that were collected using the original Sloan hardware. The \textsc{redvsblue}\footnote{\href{https://github.com/londumas/redvsblue}{https://github.com/londumas/redvsblue}} principal component analysis algorithm \citep{lyke2020ApJS..250....8L} is applied to the spectroscopic data to estimate redshifts. The eBOSS catalogues are split into the North Galactic Cap (NGC) and the South Galactic Cap (SGC), since each cap has a different targeting efficiency. The main sample spans the redshift range of $0.8<z<2.2$, and contains $218209$ QSOs in the NGC covering $2860~{\rm deg}^2$ and $125499$ QSOs in the SGC covering $1839~{\rm deg}^2$. The eBOSS program also observed 72667 Ly-$\alpha$ QSOs over $2.2<z<3.5$.
The catalogue data includes the angular position in the sky (RA, DEC) as well as the redshift Z for each galaxy. Each data point is paired with a set of weights to account for survey completeness, fiber collision pairs, and imaging systematics,
\begin{equation}\label{eq:wtot}
    w_{\rm tot} = w_{\rm systot} \times w_{\rm noz} \times w_{\rm cp} \times w_{\rm FKP},
\end{equation}
where $w_{\rm systot}$ represents imaging systematic weight, $w_{\rm noz}$ is redshift completeness weight, $w_{\rm cp}$ is close pair weight, and $w_{\rm FKP}$ is the FKP weight \citep{FKP94} defined as,
\begin{equation}
    w_{\rm FKP} = \frac{1}{1+n'_q(z) P_0},
\end{equation}
with $n'_q(z)$ being the redshift distribution of quasars and the amplitude of the power spectrum set to $P_0=6000~h^{-3}{\rm Mpc}^3$. The prime superscript $(')$ denotes the application of a completeness weight and the q subscript indicates the number density of the quasar data catalogs, rather than randoms.  The leftmost panels in Figure \ref{fig:denmaps} show the projected number density of QSOs in the NGC and SGC after undergoing this standard treatment. 
In addition to the data catalogue, a randoms catalogue is paired that matches the footprint geometry of the data.  Randoms are generated from randomly sampling the survey following the same selection criteria as the data, but with no inherent clustering signal. The randoms catalogue contains 50 times the number of objects as the data catalog.

\subsection{Systematic Error Mitigation}
As described in Eq.~\ref{eq:wtot}, each QSO and random object is weighted by the systematic weight $w_{\rm systot}$ to account for observational effects such as those caused by Galactic extinction, survey depth, and seeing. The default systematic weight $w_{\rm systot}$ is derived from a linear regression analysis minimizing the fluctuations in the mean projected number density of QSOs against a set of imaging templates \citep{Ross2020}. These templates estimate observing conditions such as:
\begin{itemize}
  \item PSF$_i$: The point spread function in the $i$-band
  \item Sky$_i$: The sky magnitude in the $i$-band
  \item EBV: Galactic extinction (reddening due to dust)
  \item depth$_g$: The image depth in the $g$-band
\end{itemize}
 
However, it has been shown in our companion paper \citep{rezaie2021} that linear regression is not able to adequately clean the data, clearly requiring nonlinear regression.  Also, an additional template for local stellar density (NStar) constructed from the Gaia spacecraft \citep{gaia2018} is required to reduce systematic uncertainties under a threshold set by systematic-free simulations. \cite{rezaie2021} develops a neural network-based approach to derive a new set of $w_{\rm systot}$, which accounts for nonlinear systematic effects.

Various NN setups are investigated to find the optimal set of imaging templates and \textsc{HEALPix} resolution. For details on the NN setups, see \cite{rezaie2021}, but a summary is provided here. Networks are divided into those which calculate weights using imaging templates in \textsc{HEALPix} with either \textsc{nside}=512 or 256. Additionally, the training of neural networks are performed with various combinations of templates for Galactic foregrounds and SDSS imaging properties. The first set includes \textit{Galactic extinction} \citep{schlegel1998maps}, \textit{neutral hydrogen column density} \citep{Hi4pi2016}, and \textit{stellar density} from the Gaia spacecraft \citep{gaia2018}. The SDSS-specific templates are \textit{seeing}, \textit{sky brightness}, and \textit{survey depth} in four bands \citep[\textit{g, r, i, z};][]{fukugita1996sloan}. The remaining two maps are \textit{run} and \textit{airmass}. To test for any redshift dependence, the neural network treatment is applied either to the entire sample covering $0.8<z<2.2$ as a whole or two redshift bins. In the second scenario, we split the main sample into $0.8<z<1.5$ and $1.5<z<2.2$, and perform the neural network training on each subsample separately. All treatments are conducted to each Galactic cap separately due to different targeting efficiencies. Table \ref{tab:NNlabels} lists the labels used throughout this paper, and their corresponding NN setup.  
\begin{table}
\centering
\caption{Labelling conventions and the corresponding NN setup. The default label indicates the linear regression used in \citet{Ross2020}. The bolded label indicates our benchmark NN setup.}

\normalsize
\resizebox{1.0\columnwidth}{!}{% use resizebox with textwidth
\begin{tabular}{ c | c  c  c}
 label & input templates & zsplit & \textsc{healpix nside}\\
 \hline
 \hline
 default &  PSF$_i$, Sky$_i$, EBV, depth$_g$   & 1  &  512 \\
 \hline
  \textbf{known-1z-H512} &  PSF$_i$, Sky$_i$, EBV, depth$_g$, NStar   & 1  &  512 \\
  known-2z-H512 &  PSF$_i$, Sky$_i$, EBV, depth$_g$, NStar   & 2  &  512 \\
  all-1z-H512 &  PSF$_i$, Sky$_i$, EBV, depth$_g$, NStar + 12 maps & 1  &  512 \\
  all-2z-H512 &  PSF$_i$, Sky$_i$, EBV, depth$_g$, NStar + 12 maps & 2  &  512 \\
  known-1z-H256 &  PSF$_i$, Sky$_i$, EBV, depth$_g$, NStar   & 1  &  256 \\
  known-2z-H256 &  PSF$_i$, Sky$_i$, EBV, depth$_g$, NStar   & 2  &  256 \\
  all-1z-H256 &  PSF$_i$, Sky$_i$, EBV, depth$_g$, NStar + 12 maps & 1  &  256 \\
  all-2z-H256 &  PSF$_i$, Sky$_i$, EBV, depth$_g$, NStar + 12 maps & 2  &  256 \\
  \hline
\end{tabular}% close resizebox
}
\label{tab:NNlabels}
\end{table}

\begin{figure*}
    \centering
    \includegraphics[width=\textwidth]{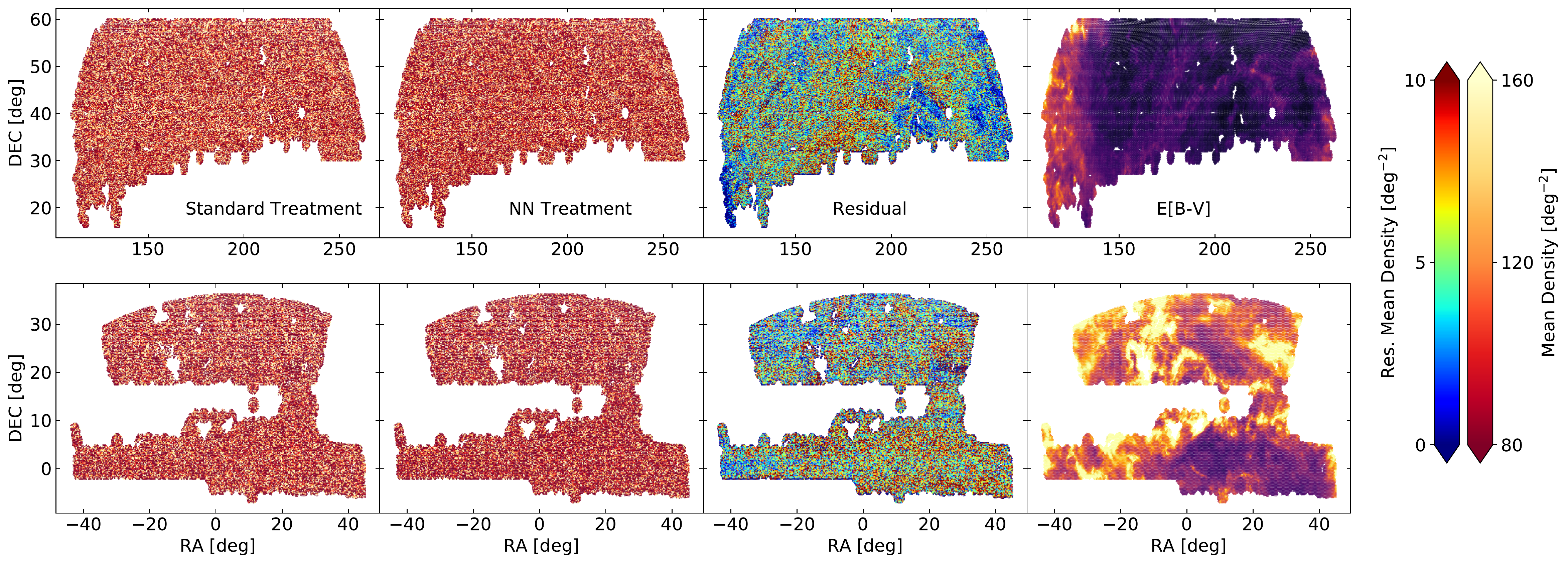}
    \caption{Projections of the number density of QSOs, treated with default systematic mitigation and NN mitigation.  Also shown is the residual density between the two, and the E(B-V) feature. The top row is a projection of the NGC and the bottom a projection of the SGC.}
    \label{fig:denmaps}
\end{figure*}

The projection maps of the systematic-corrected QSO density using the default w$_{\rm sysytot}$ weight and using the known-1z-H512 weight are shown in the left two columns of Figure \ref{fig:denmaps}. The third column shows the difference of the two treatments. Also shown as an example in the far right is a map of the E(B-V) imaging feature (extinction due to dust) from \cite{schlegel1998maps}. Even though the default treatment removes the most of the systematic effect, one can still identify the visual correlation between the difference map and the map of the E(B-V) imaging feature. \citet{rezaie2021} shows that correcting for this residual makes a large difference for removing a spurious signal in the measured power spectrum at very low $k$, and therefore neural network-based systematic weights enable the robust measurement of the primordial non-Gaussianity.
In this paper, we are testing if this residual has an impact on the BAO measurement of eBOSS QSOs.

\section{Methods}\label{sec:methods}

\subsection{Calculating the Power Spectrum}

To measure the BAO feature, statistical properties of the galaxy density field are employed. The correlation function $\xi({r})$ is a measure of the likelihood of a galaxy being found at a distance ${\textbf{r}}$ from another galaxy.  

\begin{equation}
   \xi({r}) =  \langle \delta({\textbf{x}})\delta({\textbf{x}}+{\textbf{r}})\rangle, 
    \label{eq:coorfunc}
\end{equation}
with $\delta(\textbf{r})$ as an overdensity
\begin{equation}
    \delta({\textbf{r}}) = \frac{\textit{n}({\textbf{r}})}{\overline{n}({\textbf{r}})} -1,
\end{equation} 
where \textit{n} is the number density of galaxies and $\overline{n}$ is the average number density of the sample.  

The correlation function can be thought of as the variance of the overdensity field.
The power spectrum $P({k})$ is the Fourier transform of the correlation function $\xi({r})$ and is a function of wavenumbers \linebreak $k = 2\pi/\lambda$ instead of spatial distance $r$.  

In practice, the power spectrum is calculated with the Yamamoto estimator \citep{Yamamoto06} using fast Fourier transforms as in \cite{Bianchi15}. We utilize these techniques with the method from \cite{Hand17} and the package \textsc{nbodykit}\footnote{\url{https://nbodykit.readthedocs.io/}}, an open-source python package that utilizes parallel computing \citep{Hand18}.  The multipoles of the power spectra are estimated as 
\begin{equation}
    P_\ell(k) = \frac{2\ell +1}{A} \int \frac{1}{4\pi}F_0(\textbf{k})F_\ell(\textbf{-k})d\Omega_k
\end{equation}
with
\begin{equation}
    F_\ell(\textbf{k}) = \frac{4\pi}{2\ell+1} \sum_{m=\ell-1}^{\ell}Y_{\ell m} (\textbf{\^{k}}) \int F(\textbf{r})Y_{\ell m}^{*} (\textbf{\^{r}}) e^{i\textbf{k} \cdot \textbf{r}}
\end{equation}
$F(\textbf{r})$ is the weighted density field, given as 
\begin{equation}
    F(\textbf{r})=w_{\rm FKP}(\textbf{r})[n'_q(\textbf{r})−\alpha ' n′_s(\textbf{r})],
\end{equation}
and
\begin{equation}
    A = \int d \textbf{r} [n'_q (\textbf{r}) w_{\rm FKP}(\textbf{r)}]^2 .
\end{equation}
The number density of quasars $n'_q$ and randoms $n'_s$ include completeness weights. The ratio between the two is given by $\alpha'$.

A fiducial cosmology is used to convert the catalogue coordinates of angles and redshifts to distances. We use the same cosmological parameters used in the previous analysis of the QSO sample \citep{Neveux20}.
\begin{equation}
    \begin{aligned}
    h=0.676, \quad \Omega_m =0.31, \quad \Omega_\Lambda =0.69, \\ \quad \Omega_bh^2 = 0.022, \quad \sigma_8 = 0.8
\end{aligned}
\end{equation}
This is also used to create a template power spectrum which is used in constructing the model.

\begin{figure}
    \centering
    \includegraphics[width=\columnwidth]{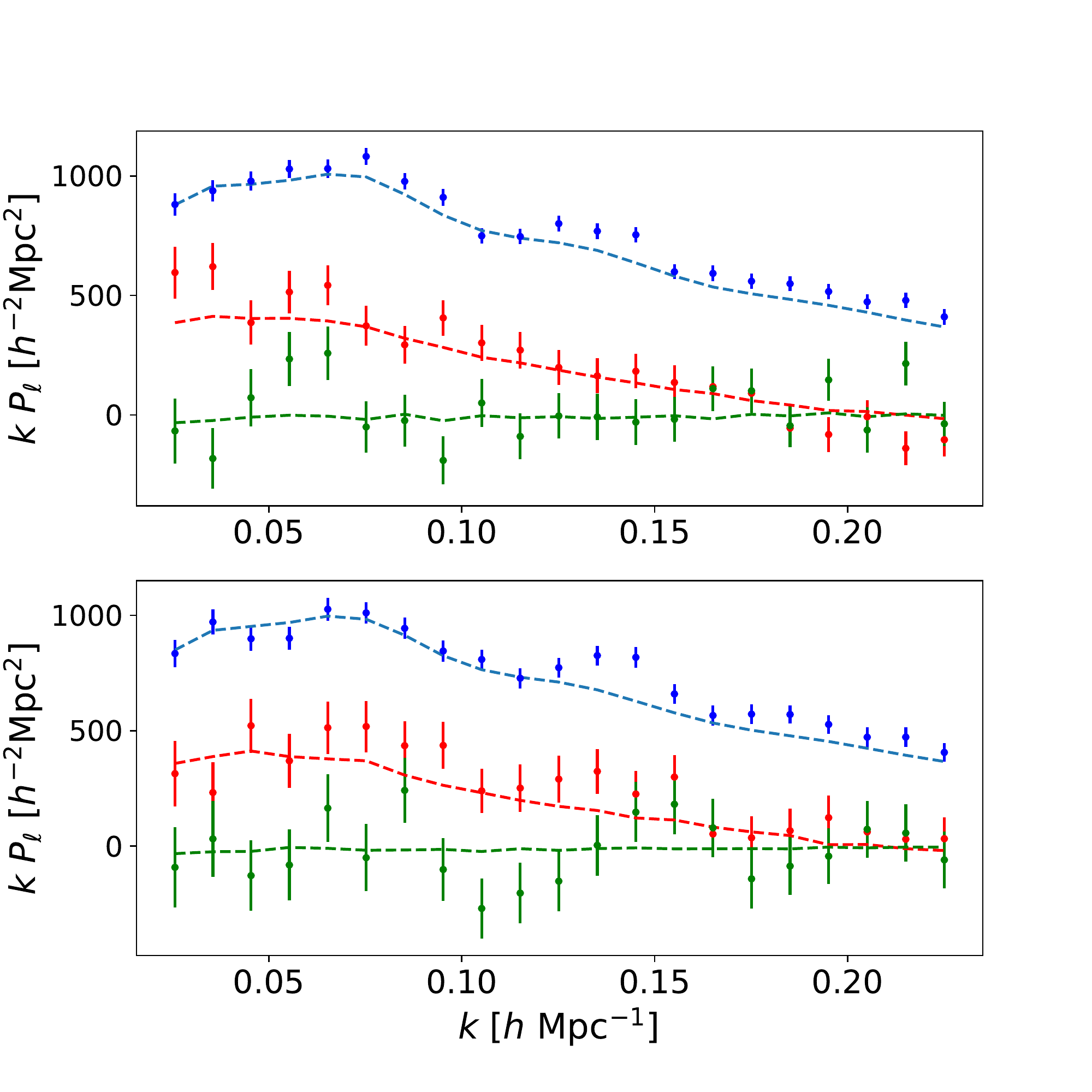}
    \caption{The monopole (blue) quadrupole (red) and hexadecapole (green) of the DR16 QSO clustering data using the default linear weight.  The NGC is shown in the top panel and the SGC in the bottom.  Error bars are derived from the covariance matrix of EZmocks.  The dashed line is the mean of the mocks.}
    \label{fig:defdatEZ}
\end{figure}

Since the overdensities must be calculated at discrete points, the algorithm must construct a mesh over the catalogue volume.  The power spectrum is calculated on a 512$^3$ mesh with a TSC mesh interpolation window. Interlacing is used to counter the effect of aliasing \citep{Interlacing}.  
 The maximum $k$ value, known as the Nyquist frequency, is then 
\begin{equation}
    k_n = \pi N_{\rm mesh}/L_{\rm box},
\end{equation}
where $L_{\rm box}$ is the length of the box that encloses the catalogue volume in $h^{-1}$Mpc.  The $k$ range for the BAO fit is well within the Nyquist frequency. The NGC and SGC QSO power spectra with the default $w_{\rm systot}$ weighting is shown in Figure \ref{fig:defdatEZ}, as well as the mean of the eBOSS EZmock power spectra that are used to calculate the covariance matrix for fitting.

\subsection{Power Spectrum Fitting}\label{subsec:fitting}
The power spectrum data needs to be compared to a model in order to extract cosmological information.  The power spectrum model used here is taken from \cite{Beutler16} and we are following \cite{Neveux20} for all fiducial conventions. For the BAO fitting model, the strategy is to single out the BAO feature from the power spectrum by marginalizing over all non-BAO features by introducing many free parameters. It first builds a power spectrum with damped BAO wiggles from a smooth model without the BAO signal.  The smooth anisotropic power spectrum is given as 
\begin{equation}
    P_{\rm sm}(k,\mu) = B^2(1+\beta\mu^2)^2P_{\rm sm,lin}(k)F_{\rm fog}(k,\mu,\Sigma_s).
    \label{eq:Psmkmu}
\end{equation}
The parameter $B$ is used to marginalize over the power spectrum amplitude, and (1+$\beta\mu^2$) accounts for RSD Kaiser effects.  In this expression, $\beta$ is defined as $f/B$, where $f$ is linear growth factor. $P_{\rm sm,lin}(k)$ is found from fitting the BAO-absent or no-wiggle $P_{\rm nw}(k)$ model of Eisenstein and Hu \cite{Eisenstein98} with the linear power spectrum template $P_{\rm lin}(k)$.  $F_{\rm fog}$ is a term for the non-linear Finger-of-God effect, and damps the BAO wiggles
\begin{equation}
    F_{\rm fog}(k,\mu,\Sigma_s) = \frac{1}{(1+k^2\mu^2\Sigma_s^2/2)^2}.
\end{equation}
with $\Sigma_s$ as a damping term due to velocities at small (nonlinear) scales.  The anisotropic power spectrum with the BAO feature is then
\begin{equation}
    P(k,\mu) = P_{\rm sm}(k,\mu) \times \big[1 + (O_{\rm lin}(k)-1)e^{[-k^2\mu^2\Sigma_\parallel^2 + k^2(1-\mu^2)\Sigma_\perp^2]/2}\big],
\end{equation}
where $\Sigma_\parallel$ and $\Sigma_{\perp}$ are anisotropic non-linear damping terms. Observed BAO wiggles are damped relative to the linear theory model due to cosmic structure formation. Following \cite{Neveux20}, $\Sigma_\parallel$, $\Sigma_\perp$ and $\Sigma_s$ are fixed to fiducial values of 8, 3, and 4 \ihMpc, respectively. $O_{\rm lin}(k)$ isolates the oscillations of the BAO signal from the fiducial linear power spectrum and is defined as 
\begin{equation}
    O_{\rm lin}(k) = \frac{P_{\rm lin}(k)}{P_{\rm sm,lin}(k)}.
\end{equation}
$P(k,\mu)$ is then projected along Legendre spherical harmonics.  The monopole ($\ell=0$), quadrupole ($\ell=2$), and hexadecapole ($\ell=4$) are 

\begin{equation}
    P_0(k) = \frac{1}{2}\int_{-1}^{1} P(k,\mu)d\mu,
\end{equation}    
\begin{equation}
    P_2(k) = \frac{5}{2}\int_{-1}^{1} P(k,\mu)\mathcal{L}_2(\mu)d\mu,
\end{equation}    
\begin{equation}
    P_4(k) = \frac{9}{2}\int_{-1}^{1} P(k,\mu)\mathcal{L}_4(\mu)d\mu,
\end{equation}
where $\mathcal{L}_i(\mu)$ is the ith order Legendre polynomial.  

To each multipole, three polynomial terms in $k$ are added to marginalize over the overall shape of the power spectrum
\begin{equation}
    A_\ell = \frac{a_{\ell,0}}{k} + a_{\ell,1} + a_{\ell,2}k.
\end{equation}
These polynomial terms are linear and as such are solved at each point in parameter space using a linear least-squares method.

The model assumes a fiducial location of the BAO wiggles in the power spectrum, and to account for differences between the data and the model, the true wave-numbers and angles k$'$ and $\mu'$ are related to the observed k and $\mu$,
\begin{equation}
    k' = \frac{k}{\alpha_\perp}\big[1+\mu^2(\frac{1}{F^2}-1)\big]^{1/2},
\end{equation}
\begin{equation}
    \mu' = \frac{\mu}{\alpha_\perp}\big[1+\mu^2(\frac{1}{F^2}-1)\big]^{-1/2}.
\end{equation}
The so-called Alcock-Paczynski (AP) parameters $\alpha_\perp$ and $\alpha_\parallel$ describe the shift in the power spectrum perpendicular to and along the line-of-sight \citep{AP79}. $F$ is $\alpha_\parallel$/$\alpha_\perp$ \citep{BPH96}. The AP parameters do not shift $P_{\rm sm,lin}(k)$ in Eq.~\ref{eq:Psmkmu}, which is different from ~\citet{Beutler16}.
Our model then has 22 free parameters: 18 polynomial coefficients (3 for each multipole in each cap) and 4 additional parameters $[B_{\rm NGC},\alpha_\parallel,\alpha_\perp,B_{\rm SGC}]$.  The final power spectra multipoles are given as
\begin{equation}\label{eq:model}
    P_\ell(k) = \frac{2\ell+1}{2\alpha_\perp^2\alpha_\parallel}\int_{-1}^{1} P\big[k'(k,\mu),\mu'(\mu)\big]\mathcal{L}_\ell(\mu)d\mu +A_\ell(k).
\end{equation}
We must also take into account the geometry of the survey, i.e. the survey window function.  We convolve the power spectrum multipoles with the survey window using matrices produced using the formalism described in \cite{Beutler19}. We obtain 
\begin{equation}
    \hat{P}_\ell(k) = \textbf{W}P_\ell(k)
\end{equation}
where W is the window function matrix from \cite{Beutler2021}.

The $\alpha$ values are the key parameters that describe the location of the BAO in the power spectrum.  They invoke the power of the BAO as a standard ruler.  The parameter $\alpha_\parallel$ gives line-of-sight information, which is related to the Hubble distance $D_H = c/H(z)$ (see \cite{Anderson14}) by 
\begin{equation}
    \alpha_\parallel = \frac{D_{\rm H}(z)/r_{\rm drag}}{D_{\rm H}^{\rm fid}(z)/r^{\rm fid}_{\rm drag}},
\end{equation}
where the $fid$ superscript indicates values calculated using a fiducial model.  Similarly, $\alpha_\bot$ is related to the angular diameter distance D$_M$ by 
\begin{equation}
    \alpha_\bot = \frac{D_{\rm M}(z)/r_{\rm drag}}{D_{\rm M}^{\rm fid}(z)/r^{\rm fid}_{\rm drag}}.
\end{equation}
Here $r_{\rm drag}$ is the sound horizon scale at the drag epoch.

All the other parameters are free parameters to remove non-BAO features. Additionally, $\alpha_\perp$ and $\alpha_\parallel$ can be combined as 
\begin{equation}
    \alpha_{\rm iso} = \alpha_\parallel^{1/3}\alpha_\perp^{2/3}
\end{equation}
into one $\alpha$ parameter which gives information related to the volume-averaged distance $D_v(z)$.

For the default analysis, we utilize a covariance matrix derived from 1000 synthetic catalogs \citep{EZmock20}. These mocks are generated using the effective Zel'dovich approximation \citep{Zel70} and the algorithm is able to reproduce the clustering statistics to within 1\% of N-body simulations  on scales that encompass the BAO signal \citep{Chuang15}. The creation of the mocks is described in detail in \cite{EZmock20}. In order to account for systematics, these effective Zel'dovich (EZ)mocks undergo a systematics treatment and correction with the standard linear regression.  This set of contaminated and corrected mocks (and the covariance and weights derived from them) will be referred to with the label `default'.  Uncontaminated mocks will be referred to as `null' mocks.  We also calculate the covariance matrix after applying NN weights to contaminated mocks instead of default weights and compare the results of fitting with the different covariance matrices in Section \ref{subsec:errorprop}.  In order to account for a limited number of mocks, we correct the covariance with the factor described in \cite{Hartlap07} such that the new corrected covariance matrix is 
\begin{equation}
    C^{-1} = \frac{N-n-1}{N-1}C^{-1}
\end{equation}
where $N$ in the number of mocks and $n$ is the number of data points.
We also apply a correction factor to parameter uncertainties as following \cite{Percival14}. This is 
\begin{equation}
    M_1 = \frac{1+B(n_b−n_p)}{1+A+B(n_p+1)}
\end{equation}
where $A=2/[(n_{\rm s}−n_{\rm b}−1)(n_{\rm s}−n_{\rm b}−4)]$, $B=(n_{\rm s}−n_{\rm b}−2)/[(n_{\rm s}−n_{\rm b}−1)(n_{\rm s}−n_{\rm b}−4)]$. The parameter $n_{\rm s}$ is the number of mocks used to make the covariance and $n_{\rm b}$ is the number of data points.
To obtain best-fit parameters, we maximize the likelihood function 
\begin{equation}
    L \propto e^{-(v^{T}C^{-1}v)/2}
\end{equation}
where $v = P_{\rm data} - P_{\rm model}$. We use the python package \texttt{emcee} \citep{Emcee13} to explore the parameter space with Monte Carlo Markov Chains and marginalize over parameters that are not important to the BAO signal. 
\begin{figure}
     %\begin{subfigure}
         \centering
         \includegraphics[width=\columnwidth]{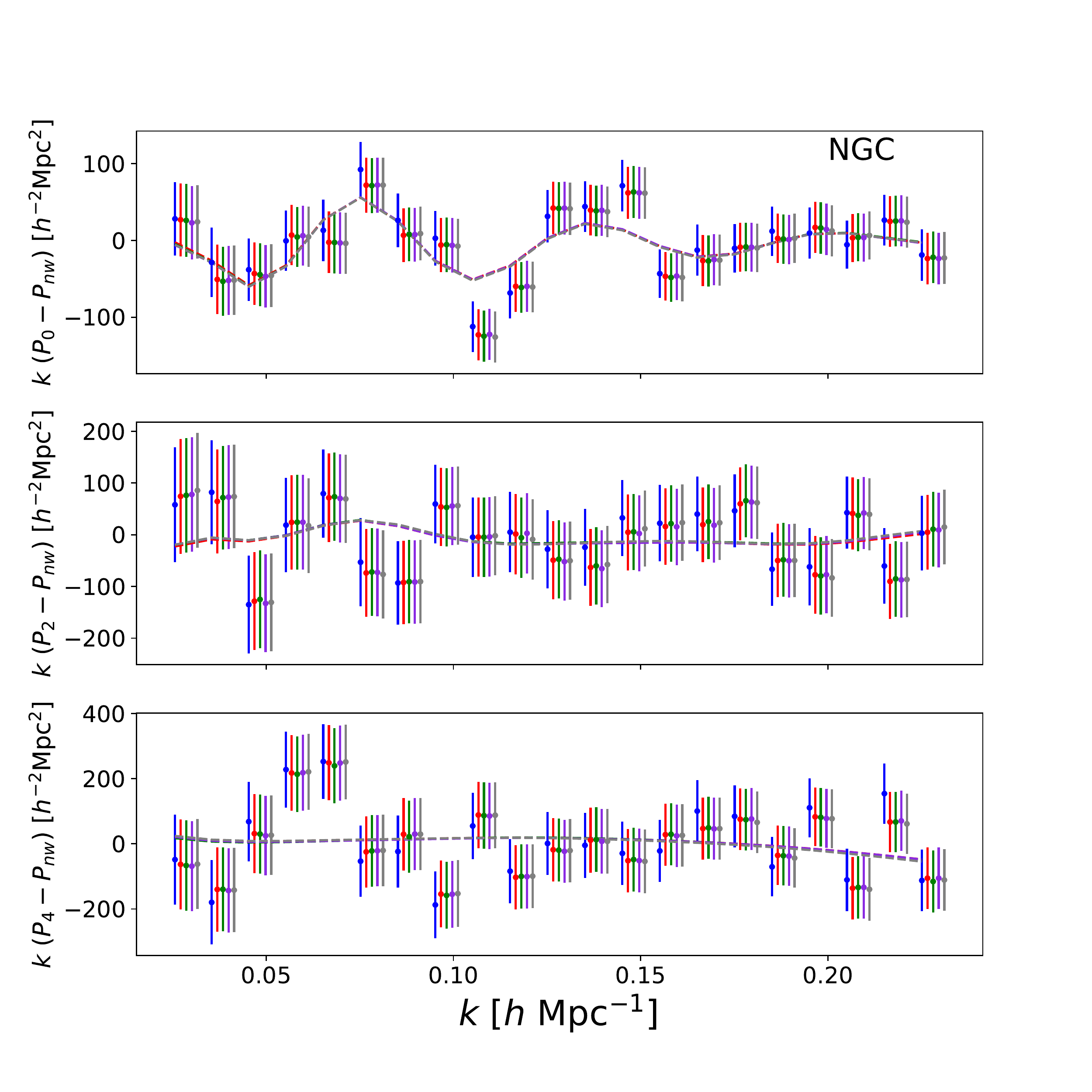}
     %\end{subfigure}
     %\begin{subfigure}
         %\centering
         \includegraphics[width=\columnwidth]{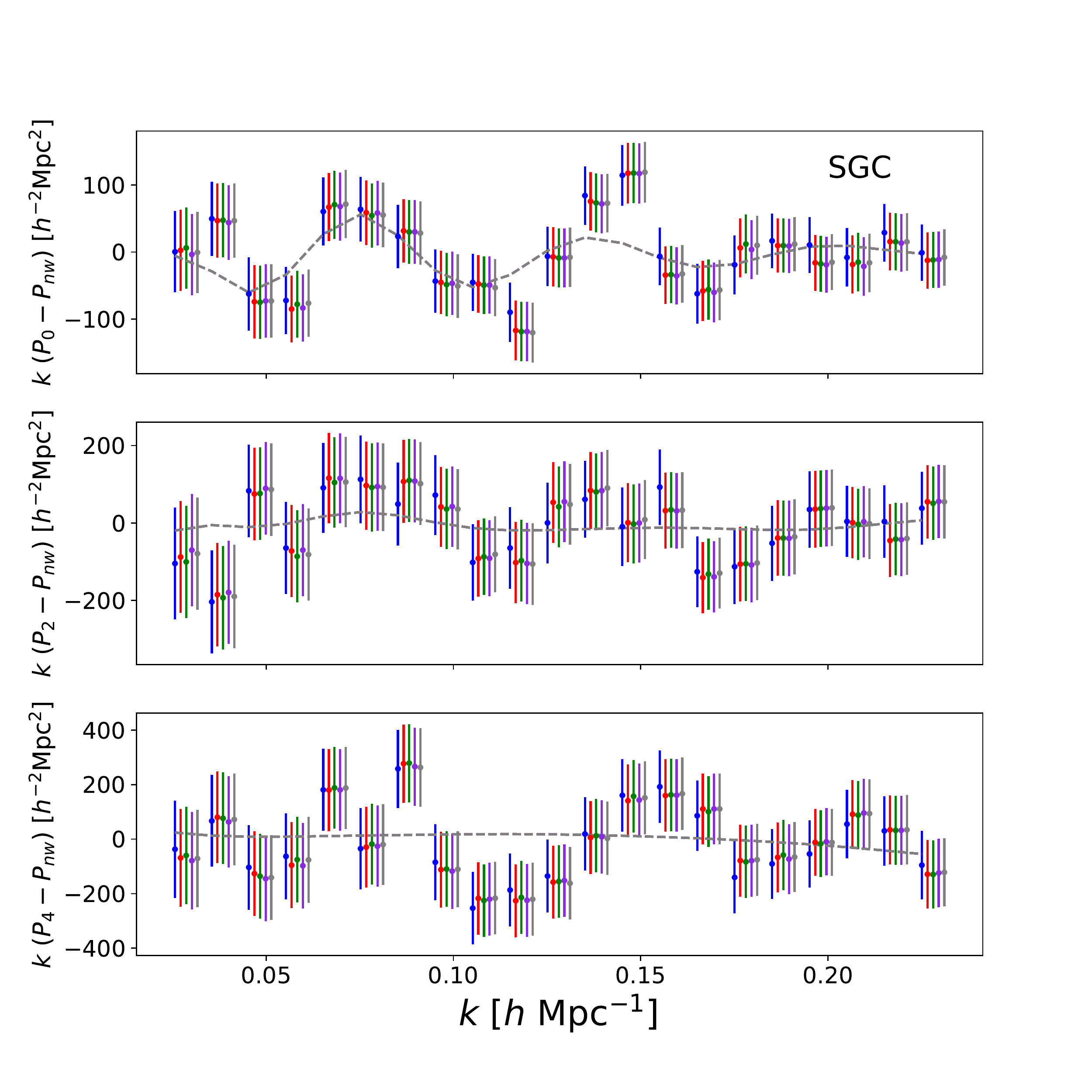}
     %\end{subfigure}
     \caption{The effect of the NN weights in the power spectrum in comparison to the default weight (blue),
     as a function of various setups of the NN method (Table \ref{tab:NNlabels}). We show the isolated BAO wiggles in the power spectra of the H512 catalogues. The first set of panels shows the BAO signal in the NGC and the second set shows the SGC.  Points are the data with error bars derived from the covariance matrix and the solid lines are the best-fit models. The wiggles are isolated by subtracting the power spectra by a smoothed spectra, which is produced from fitting the default catalogue to a smoothed model.  Data points are shifted horizontally to prevent overlap, but best-fits are not shifted. Colors correspond to the catalogues with different systematic weights: red green, purple and grey representing known-1z, known-2z, all-1z, and all-2z, respectively.}
     \label{fig:BAOwig512}
\end{figure}

\section{Results}\label{sec:results}

\subsection{Effect on the power spectrum}
Figure \ref{fig:BAOwig512} shows the effect of the NN weights in the power spectrum as a function of various setups of the NN method (Table \ref{tab:NNlabels}), in comparison to the default linear regression weight (blue). We are showing only the cases with the \textsc{HEALPix} resolution of 512 as an example. In the figure, the BAO feature is isolated by subtracting a smooth fit power spectrum from the best-fit spectrum and the error bars are taken from the covariance matrix of EZmocks treated with the default linear regression weight. Overall, we see a small difference; a few points are shifted by about half $\sigma$ at for various multipoles, and there is one point in the NGC hexadecapole that is shifted by about 1$\sigma$. The weights derived using the 256 \textsc{HEALPix} show a similar trend (included in Appendix \ref{App:256}).
In order to have a more quantitative measure of the effect on the BAO constraint, we propagate this difference in the power spectrum to the BAO measurement in the next section. Then, in~\S~\ref{subsec:errorprop} we include the effect of the NN weight on the covariance matrix.

 \renewcommand*{\thefootnote}{\fnsymbol{footnote}}

\begin{table*}

      \caption{The BAO constraints from the eBOSS QSO catalog using different mitigation methods. `Default' here presents our reproduction of the \citet{Neveux20} using exactly the same setup. 
      Posterior values are taken as the mean and standard deviation of the MCMC chains.  Shifts $\Delta\alpha$ are compared to the default fit. Catalogs are grouped by the HEALPIX resolution used to determine NN weights.  Values in parentheses are derived from chi-square minimization rather than the posterior mean.}
      \label{tab:alphatable}

    \begin{threeparttable}
        \begin{adjustbox}{max width=\textwidth}

         \begin{tabular}{ l | l  l  l  l  l  l  l  l}%{\textwidth}
 & $\alpha_\parallel$ & $\Delta \alpha_\parallel $ & $\alpha_\perp$ &  $\Delta \alpha_\perp$ & $r_{\rm off}$  &$\alpha_{\rm iso}$ & $\Delta\alpha_{\rm iso}$  &$\chi^2$\\
 \hline
   Neveux 2020\tnote{*} & 1.035 $\pm$ 0.045 & -- & 1.017 $\pm$ 0.029 & -- & --  & --- & --- & 87.63 / 104 \\
  \hline
  default & 1.0386 (1.0361) $\pm$ 0.0528 & -- & 1.0169(1.0141) $\pm$ 0.0328 & -- & -0.3923 & 1.0236 (1.0214) $\pm$ 0.0215 & -- & (88.910) / 104 \\
  \hline
  no wsys & 1.0324 (1.0303) $\pm$ 0.0524 & -0.0062 (-0.0058) & 1.0165 (1.0131) $\pm$ 0.0343 & -0.0004 (-0.0009) & -0.3828 & 1.0212 (1.0188) $\pm$ 0.0224 & -0.0024  (-0.0025) & (90.823)/104\\
  \hline
  H512 & & & & &\\
  known-1z & 1.0307 (1.0273) $\pm$ 0.0550 & -0.0079 (-0.0089) &  1.0139 (1.0110) $\pm$ 0.0348 &  -0.0030 (-0.0030) & -0.3890 & 1.0189 (1.0164) $\pm$ 0.0228  & -0.0047 (-0.0050) & (91.082)/104 \\
  known-1z\tnote{\textdagger} & 1.0292 (1.0255) $\pm$ 0.0532 & -0.0095 (-0.0107) &  1.0166 (1.0139) $\pm$ 0.0342 &  -0.0003 (-0.0002) & -0.3977 & 1.0202 (1.0177) $\pm$ 0.0222 & -0.0034 (-0.0036) & (92.861)/104 \\
  known-2z & 1.0322 (1.0284) $\pm$ 0.0545 & -0.0064 (-0.0077) & 1.0130 (1.0102) $\pm$ 0.0348 & -0.0039 (-0.0039) & -0.3716 & 1.0188 (1.0162) $\pm$ 0.0230  & -0.0048 (-0.0051)  & (91.004)/104 \\
  all-1z & 1.0321 (1.0283) $\pm$ 0.0538 & -0.0066 (-0.0078) & 1.0125 (1.0113) $\pm$ 0.0342  & -0.0044 (-0.0027) & -0.3547 & 1.0184 (1.0169) $\pm$ 0.0230 & -0.0051 (-0.0044)  &  (91.011)/104 \\
  all-2z & 1.0361 (1.0323) $\pm$ 0.0544 & -0.0025 (-0.0038) & 1.0108 (1.0083) $\pm$ 0.0339 & -0.0061 (-0.0057) &  -0.4065 &  1.0186 (1.0163) $\pm$ 0.0220 & -0.0050 (-0.0051) & (91.178)/104 \\
  \hline
  H256 & & & & &\\
  known-1z & 1.0318 (1.0282) $\pm$ 0.0560 & -0.0068  (-0.0079) &  1.0144 (1.0108) $\pm$ 0.0343 &  -0.0025 (-0.0033) & -0.3805 & 1.0196 (1.0165) $\pm$ 0.0229 & -0.0040 (-0.0048) & (92.571)/104 \\
  known-2z & 1.0310 (1.0281)  $\pm$ 0.0539 & -0.0076 (-0.0080) & 1.0138 (1.0103)  $\pm$ 0.0345 & -0.0031 (-0.0038) &  -0.4008  & 1.0189  (1.0162) $\pm$ 0.0223 & -0.0046 (-0.0052) &  (90.247)/104 \\
  all-1z & 1.0341 (1.0302) $\pm$ 0.0540 & -0.0045 (-0.0059) &  1.0133 (1.0109) $\pm$ 0.0343 & -0.0036 (-0.0032) & -0.3627 & 1.0196 (1.0173) $\pm$ 0.0229 & -0.0039 (-0.0041) &  (91.094)/104 \\
  all-2z & 1.0349 (1.0304) $\pm$ 0.0537 & -0.0038 (-0.0057) & 1.0113 (1.0095) $\pm$ 0.0347 & -0.0055 (-0.0045) &  -0.3874 & 1.0186 (1.0164) $\pm$ 0.0226 & -0.0050 (-0.0049)  & (91.749) /104 \\
\hline
    \end{tabular}
    \end{adjustbox}
    \begin{tablenotes}
    \item[*] Error is derived from the $\Delta\chi^2=1$ abscissa
    \item[\textdagger] Fit performed using a covariance matrix derived from mocks with the same NN mitigation
    \end{tablenotes}
    \end{threeparttable}

\end{table*}

\subsection{The BAO constraint}
\begin{figure}
    \centering
    \includegraphics[width=\columnwidth]{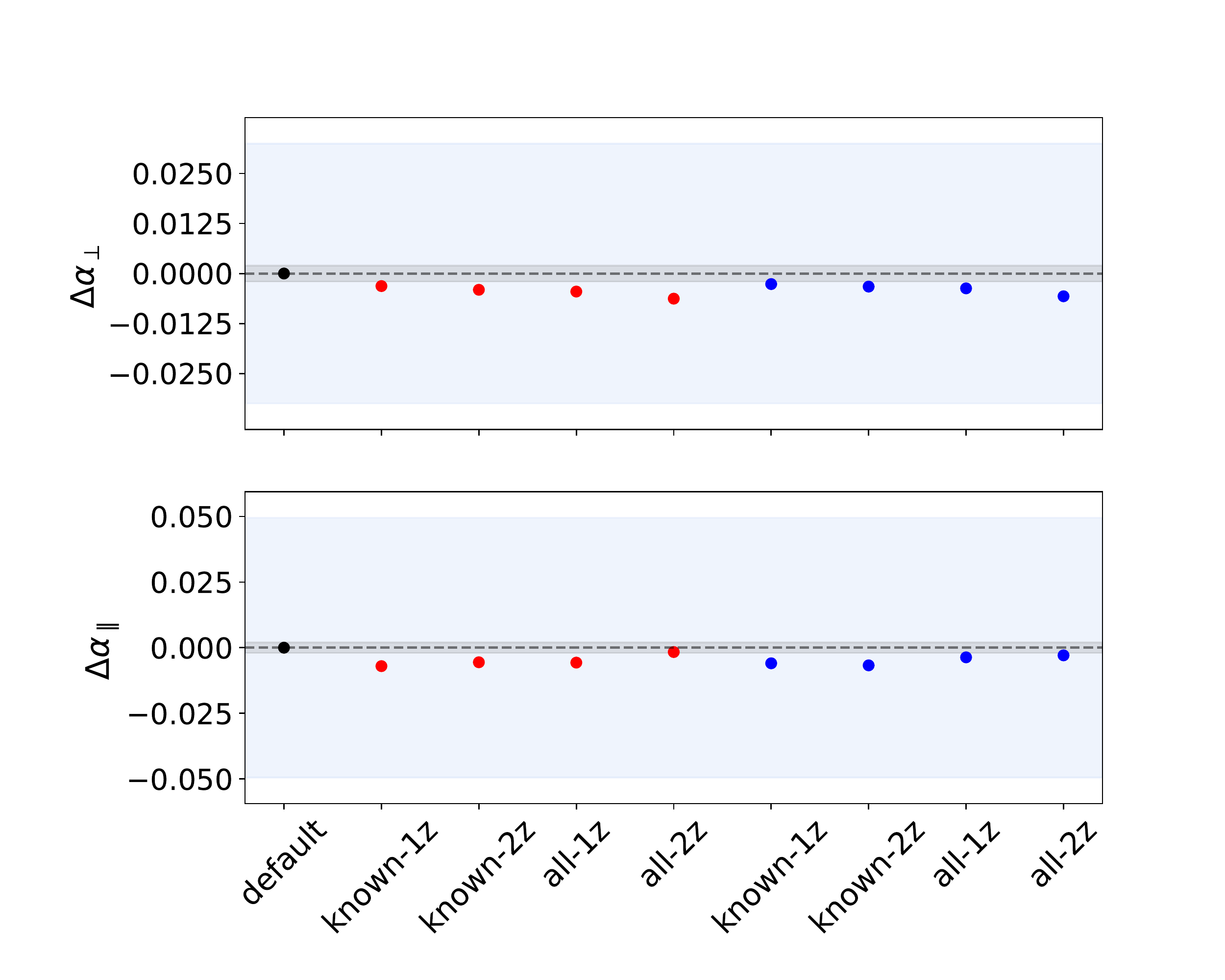}
    \includegraphics[width=\columnwidth]{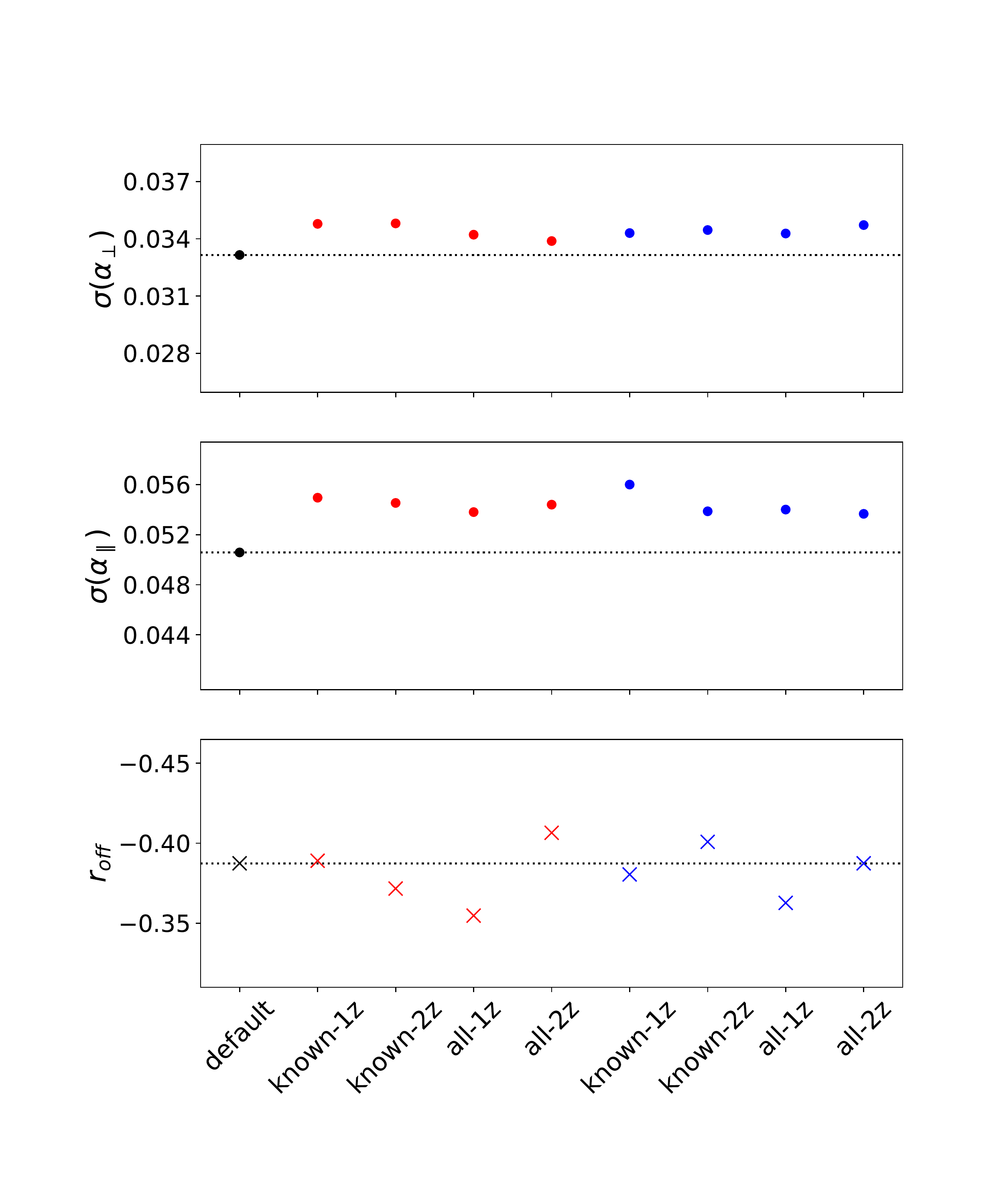}
    \caption{The best-fit BAO constraints derived from the different weights of the eBOSS QSO catalogue. We use the default covariance matrix derived using the default weight for all cases. Fits were done to the NGC and SGC simultaneously.  The points and are taken from the posterior MCMC distribution, and the blue shade represents the error of the default fit derived from the posterior. In the upper panels, the gray bar centered on the default fit corresponds to a precision of 0.2\%, which is approximately an aggregate precision of the DESI~\citep{DESIFisher,DESI16}. Red points are for H512 catalogues and blue for H256 catalogues}
    \label{fig:alphadiff}
\end{figure}

Figure \ref{fig:alphadiff} compares the best-fit 1$\sigma$ constraints on the BAO derived from the different weights. We use the default covariance matrix derived using the default weight for all cases, in order to separate the effect on the covariance matrix from the effect on the power spectrum measurement. 
Red points represent catalogues with NN weights derived from a 512 \textsc{HEALPix} mesh resolution, and blue points for a 256 mesh, in comparison to the default (black).
Fits are simultaneously done to the NGC and SGC data, which are considered separate data sets. The blue shade shows the error from the default fit, and the gray shade shows the precision of 0.2\% which is comparable to the expected aggregate precision of the entire DESI survey~\citep{DESIFisher,DESI16}. 

Table \ref{tab:alphatable} shows the corresponding values of the BAO constraint for each case of Figure \ref{fig:alphadiff}, as well as the result from \cite{Neveux20}.  The `default' here is our reproduction of \cite{Neveux20}. Our error is 10-18\% bigger as we are deriving errors based on the standard deviation from the MCMC chains while the reported constraint in \cite{Neveux20} is based on $\Delta \chi^2=1$ abscissa. Throughout this paper, we consistently use the standard deviation of the MCMC chain as our error estimates. Except for `known-1z{\textdagger}', all cases use the default covariance matrix.

\begin{figure}
     %\begin{subfigure}
         \centering
         \includegraphics[width=\columnwidth]{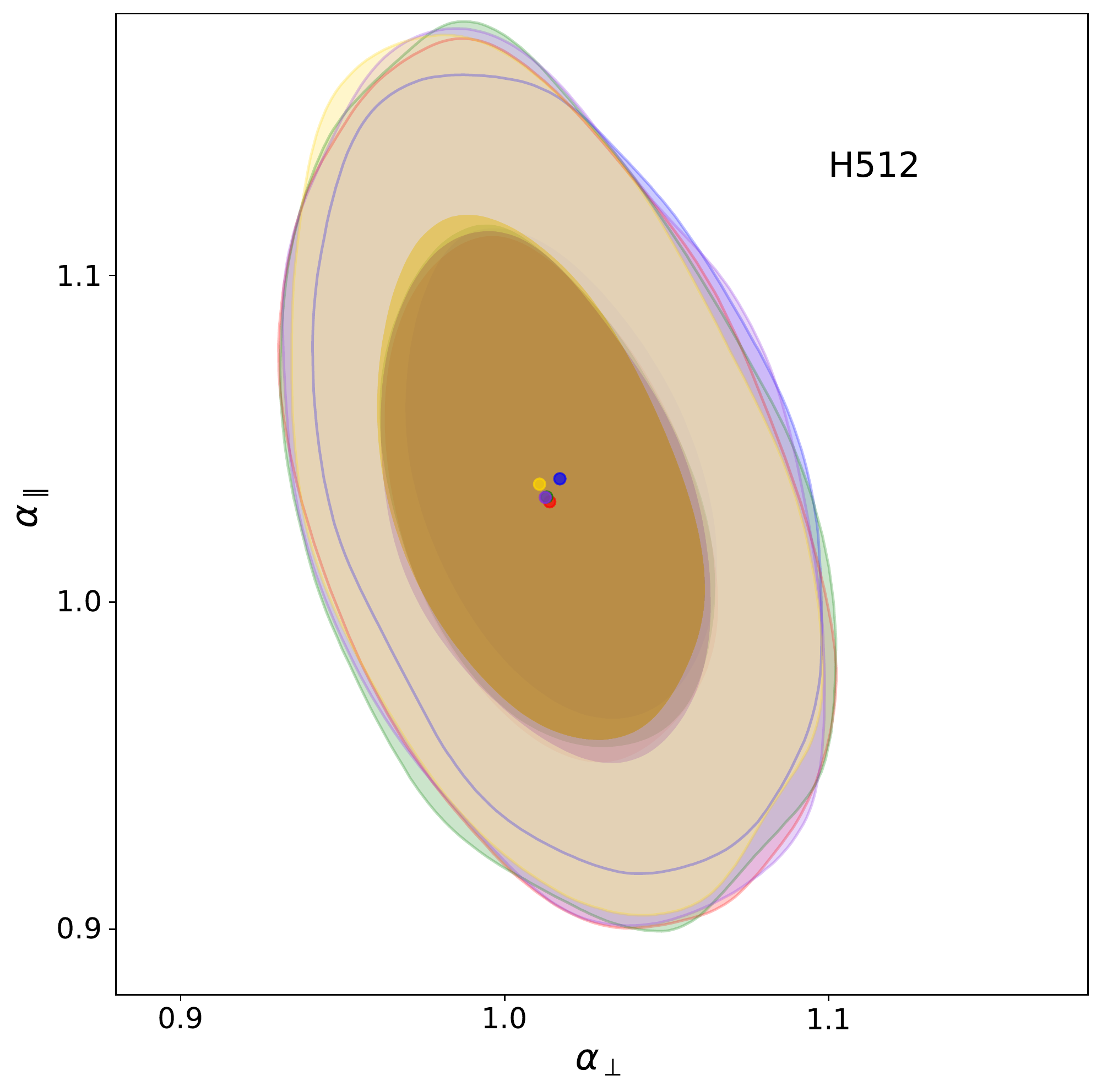}
     %\end{subfigure}
     %\begin{subfigure}
        %\centering
         \includegraphics[width=\columnwidth]{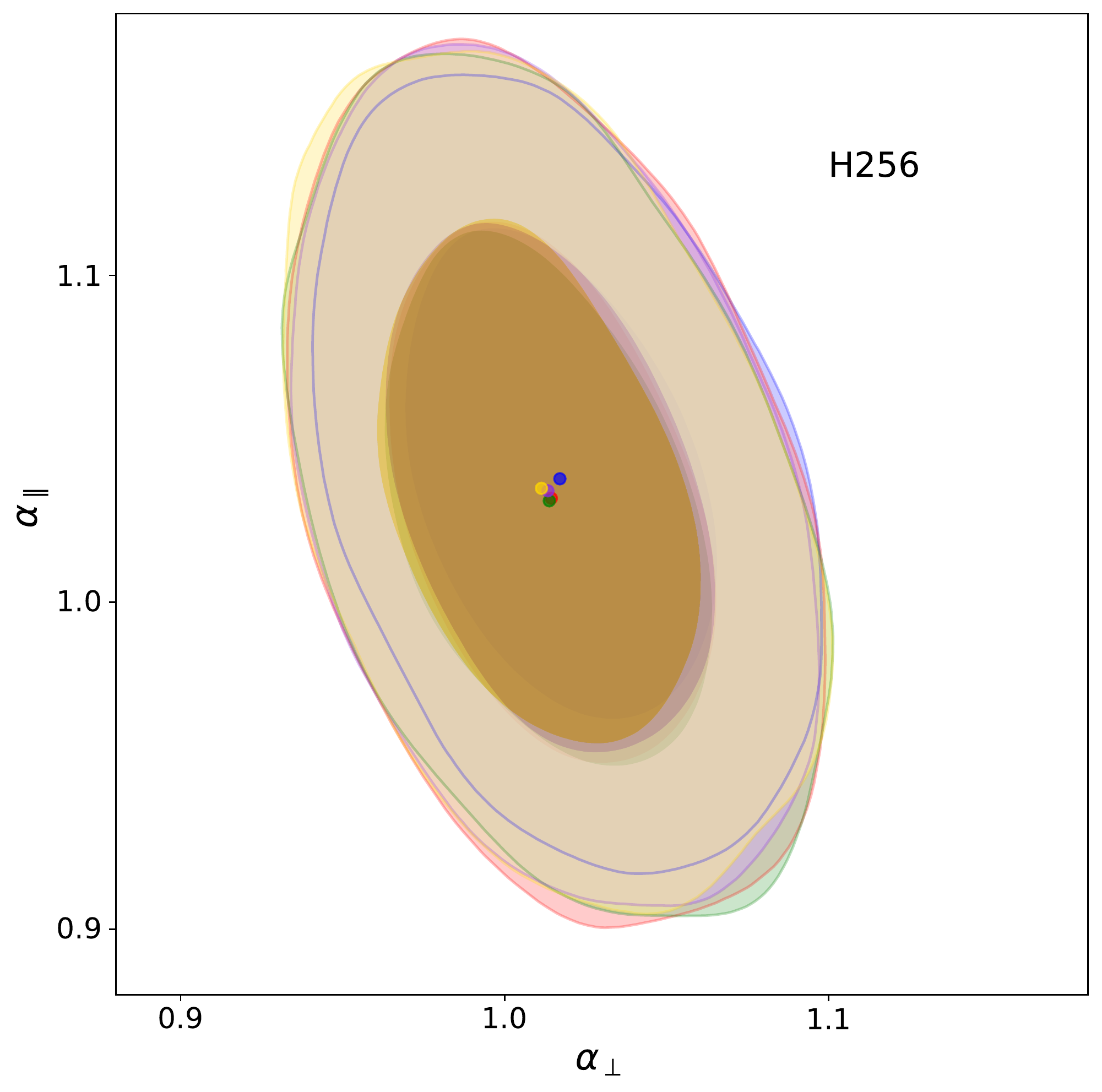}
     %\end{subfigure}

     \caption{1 and 2$\sigma$ error ellipses of the simultaneous NGC+SGC fits, corresponding Fig.~\ref{fig:alphadiff}. The top panel is for catalogues that use a 512 \textsc{HEALPix} resolution and the bottom for those with a 256 resolution.  The colors are blue (default), red (known-1z), green (known-2z), violet (all-1z) and gold (all-2z).\FB{Since there is noise in theses MCMC chains, have you checked that any difference you find is significant? If you split the chain randomly into two chains and derive the same parameters, is the noise in those parameters smaller than the difference you find between these chains?}}
     \label{fig:alpha_ellipse}
\end{figure}

We find that the offset due to the mitigation method is small, less than 0.6\% for $\aperp$ and less than 1\% for $\apar$. This is at most $0.2\sigma$ in terms of the final statistical precision reported; therefore, we consider the bias due to observational systematics fairly insignificant. While the offsets are consistent within the statistical error, we can make a few statements about the behavior of the BAO signal when different systematics mitigations are used.  There are consistently negative, meaning that the best fit BAO scales shift toward to a smaller scale with a more efficient mitigation.  We find that the shift in the isotropic scale $\alpha_{\rm iso}$ , which is calculated from $[\aperp^2\apar]^{1/3}$, shows more consistent deviation of 0.4-0.5\% from the default mitigation, compared to the anisotropic $\alpha_{\parallel}$ and $\alpha_\perp$ measurements. In terms of the cross-correlation coefficients, we are recovering values of around $-0.4$ which is expected for a BAO only analysis~\citep{SeoFisher2007}, except for `all-1z'. We do not see a clear indication of the goodness of the fit ($\chi2$ divided by the degrees of freedom) depending on the mitigation schemes. Also, redshift-dependent mitigation (i.e., `2z') does not show an advantage. The 2D distributions of the MCMC chains corresponding to fits with different weights are shown in Figure \ref{fig:alpha_ellipse}, with  68\% and 95\% confidence intervals. All NN weights are closely clustered with respect to the default (blue) and more consistent along the diagonal line between opposite corners, i.e., a more consistent $\alpha_{\rm iso}$.

To test the level of noise in the chains, we compare the difference due to the mitigation methods with a typical fluctuation due to the convergence of the chains. We randomly split the chains into two subsets and find that the average difference in the mean $\alpha$s between the two splits is at the level of 30\% of the difference we observe between the default and our error-propagated fit (`known-1z\textdagger' in Section \ref{subsec:errorprop} and Table \ref{tab:alphatable}). I.e. the noise due to convergence is smaller than the difference we report here.

\begin{table*}
\centering
\caption{The BAO constraints from the eBOSS QSO catalog using the default and the NN mitigation methods when the BAO fitting parameters are varied. As in Table \ref{tab:alphatable}, `default' presents our reproduction of the \citet{Neveux20} using exactly the same setup. 
%Results of the fitting the catalogs with a slightly different model.  
Listed are fits of the default catalog with default covariance, and the known-512-1z catalog with covariance derived from the same NN setup.  Shifts $\Delta\alpha$ in the default rows are compared to the default fit with the standard model.  Shifts $\Delta\alpha$ in the `known' rows are compared to the default fit with the same model.  We compare a model with $f$ as a free parameter and a model with an additional $k^{-2}$ polynomial. }
\normalsize
\resizebox{\textwidth}{!}{% use resizebox with textwidth
\begin{tabular}{ l | l  l  l  l  l  l  l  l}
 & $\alpha_\parallel$ & $\Delta \alpha_\parallel$ & $\alpha_\perp$ & $\Delta \alpha_\perp$ & $r_{\rm off}$ & $\alpha_{\rm iso}$ & $\Delta\alpha_{\rm iso}$ & $\chi^2$ / dof\\
  \hline
   Neveux 2020 & (1.035) $\pm$ 0.045 & -- & (1.017) $\pm$ 0.029 & -- & -- & -- & -- & 87.63 / 104 \\
  \hline
  default & 1.0386 (1.0361) $\pm$ 0.0528 & -- & 1.0169(1.0141) $\pm$ 0.0328 & -- & -0.3923 & 1.0236 $\pm$ 0.0215 & -- & (88.910)/104 \\
  default (free f) & 1.0465 (1.0493) $\pm$  0.0500 & 0.0087 (0.0132) & 1.0112 (1.0024) $\pm$ 0.0343 & -0.0058 (-0.0117) &  -0.3700 & 1.0223 (1.0178) $\pm$ 0.0223 & -0.0012 (-0.0036) & (87.266) / 103 \\
  default ($k^{-2}$) & 1.0320 (1.0295) $\pm$  0.0458 & -0.0058 (-0.0066) & 1.0182 (1.0160) $\pm$ 0.0281 & 0.0011 (0.0019) &  -0.3870 & 1.0223 (1.0204) $\pm$ 0.0186 & -0.0012 (-0.0009) & (80.815) / 98 \\
  default (no $\ell=4$) & 1.0457 (1.0415) $\pm$ 0.0545 & 0.0070 (0.0053) & 1.0181 (1.0154) $\pm$ 0.0341 & 0.0012 (0.0013) & -0.4167 & 1.0266 (1.0240) $\pm$ 0.0219 & 0.0031 (0.0027) & (47.346) / 68 \\
  \hline
   known & 1.0292 (1.0255) $\pm$ 0.0532 & -0.0095 (-0.0107) & 1.0166 (1.0139) $\pm$ 0.0342 & -0.0003 (-0.0002) & -0.3977 & 1.0202 (1.0177) $\pm$ 0.0222 & -0.0034 (-0.0036)  & (92.861) / 104\\
  known (free f) & 1.0408 (1.0377) $\pm$ 0.0494 & -0.0057 (-0.0116) & 1.0097 (1.0031) $\pm$ 0.0338 & -0.0015 (0.0008) & -0.3706 & 1.0195 (1.0145) $\pm$ 0.0220 & -0.0029 (-0.0033) & (89.572) / 103\\
  known ($k^{-2}$) & 1.0204 (1.0180) $\pm$  0.0436 & -0.0115 (-0.0115) & 1.0188 (1.0171) $\pm$ 0.0287 & 0.0006 (0.0011) &  -0.3656 & 1.0189 (1.0174) $\pm$ 0.0189 & -0.0034 (-0.0031)  & (82.044) / 98 \\
  known (no $\ell$=4) & 1.0321 (1.0272) $\pm$  0.0567 & -0.0135 (-0.0143) & 1.0183 (1.0150) $\pm$ 0.0347 & 0.0002 (-0.0004) &  -0.4253 & 1.0223 (1.0190) $\pm$ 0.0223 & -0.0044 (-0.0050) & (56.351) / 68 \\\hline
\end{tabular}
}
\label{tab:modeltable}
\end{table*}

\subsection{Error propagation}\label{subsec:errorprop}
We inspect the effect of propagating errors/covariance structure that may have been changed in the process of NN-based mitigation. We derive a self-consistent covariance matrix for `known-1z-H512', which is our benchmark NN mitigation, by applying the given method on the 1000 contaminated EZ mocks, calculating power spectra, and obtaining the covariance matrices.  We then repeat the fitting process.

\begin{figure}
    %\begin{subfigure}

    \centering
    \includegraphics[width=\columnwidth]{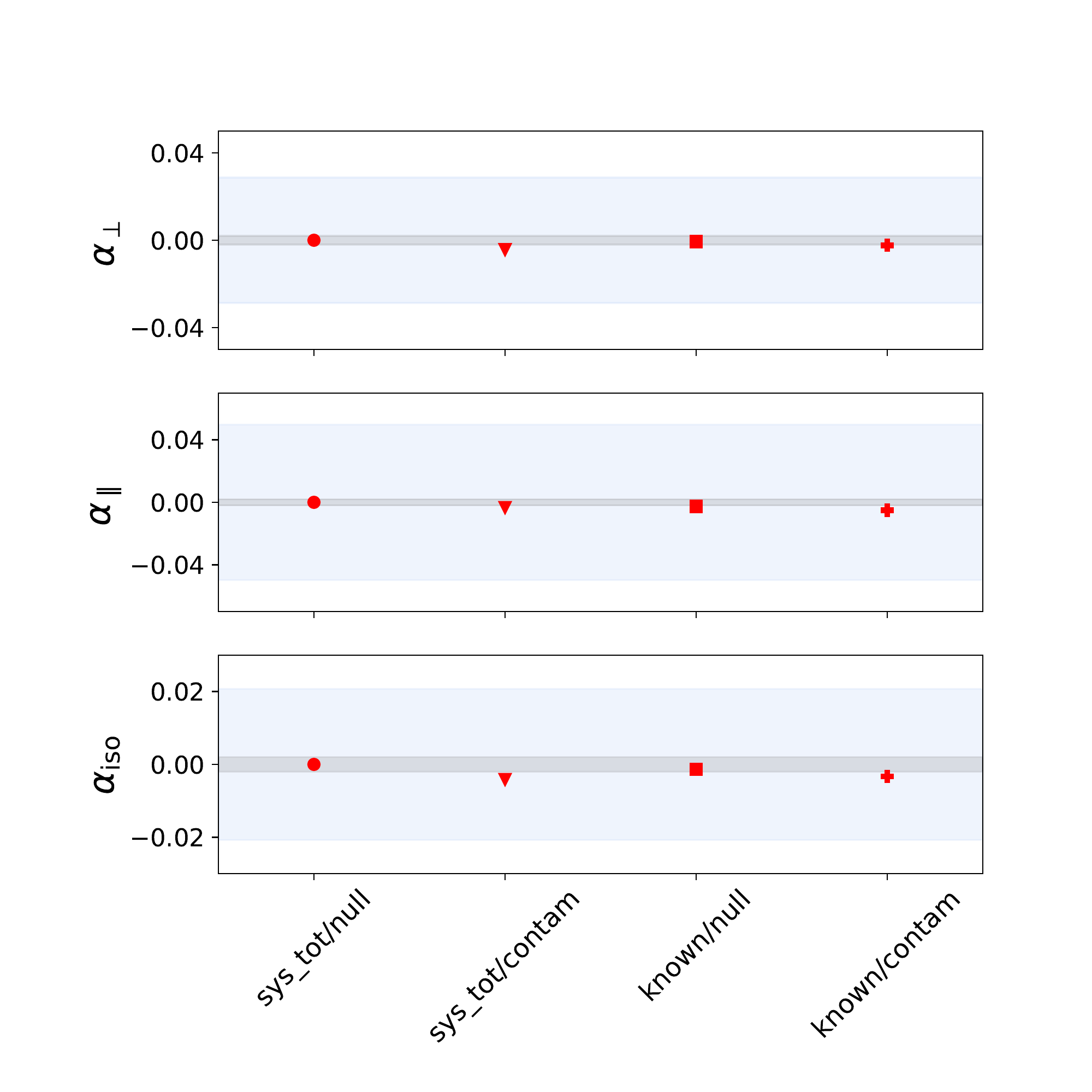}
    
    %\end{subfigure}

    %\begin{subfigure}
     %\centering
    \includegraphics[width=\columnwidth]{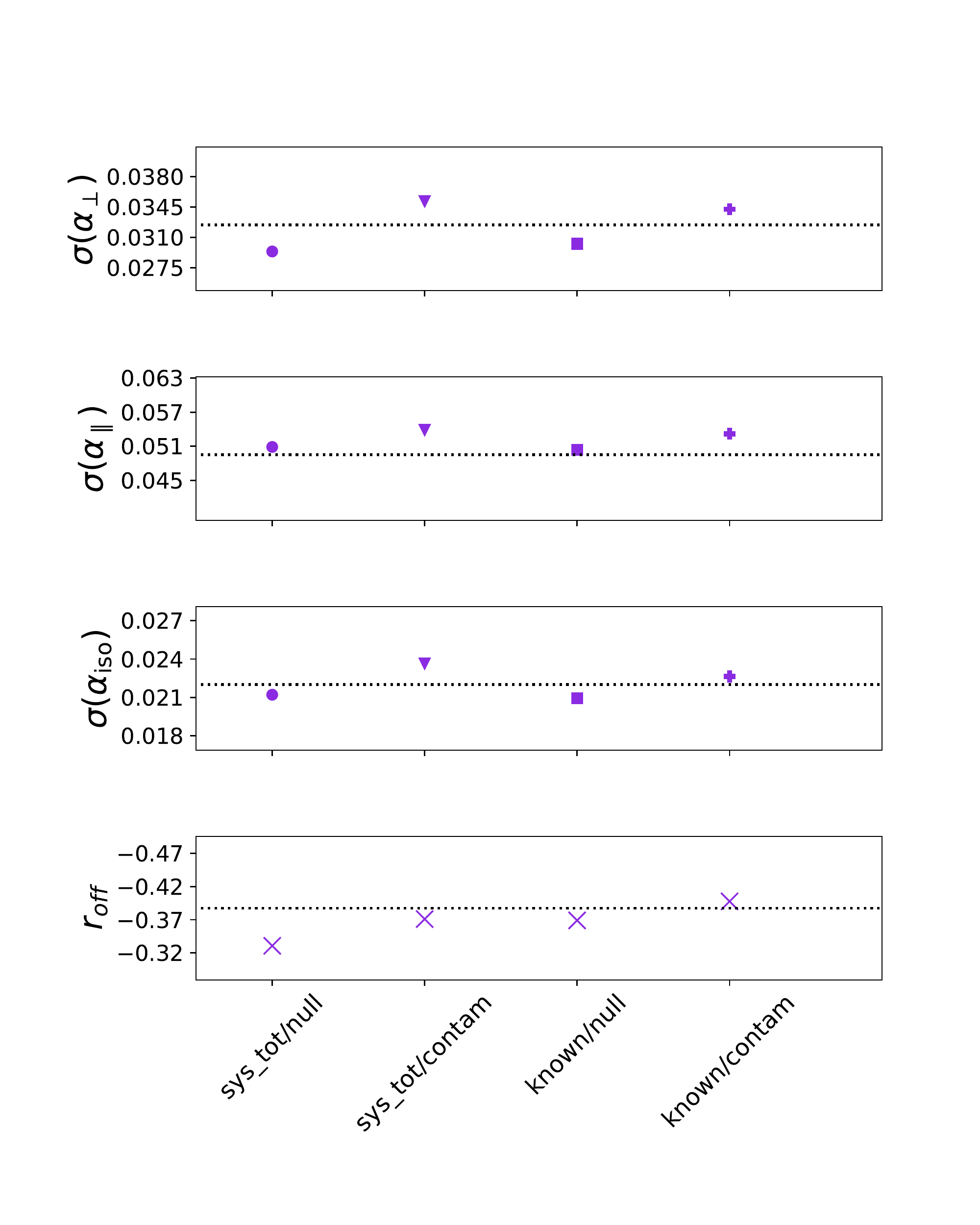}

    %\end{subfigure}

    \caption{Differences in best-fit $\alpha$ values fitted to the data with different covariance matrices \FB{Are you plotting differences? It seems you just plot alpha's? I think it would make sense to plot differences though for all plots which currently show alpha's... the variations are just too small to see and if you would plot differences these plots would be much more useful.}.  Points are the fitted values taken from MCMC posterior distributions.  All fits are done on the ``known-1z-H512" catalogue. Fits vary by using different covariances derived from null or contaminated mocks with NN of default weights.  ``Sys\_tot/null" signifies a fit using a covariance derived from null mocks with no weight (i.e. ground truth mocks).  The ``sys\_tot/contam" fit uses a covariance derived from contaminated mocks treated with the default (linear regression) weight.  The ``known/null" mocks are null mocks treated with the known NN weight and the ``known/contam" mocks are contaminated mocks that have been mitigated with the known NN weight.  The bottom panel displays the errors on $\alpha$ and the cross-correlation coefficient of each fit.}
    \label{fig:error_cov_diff}

\end{figure}

Figure \ref{fig:error_cov_diff} shows the effect on BAO constraints as we change the choice of the covariance matrix when the data vector is fixed to be the "known-1z-H512" catalogue.  
"Sys\_tot/contam" is the label for using the aforementioned default eBOSS QSO DR16 covariance matrix, as in Fig~\ref{fig:alphadiff}. 
 "Sys\_tot/null" is our designation for using the covariance matrix with no systematic effects, "known/null" is for using the covariance derived by applying the known-1z-H512 mitigation on the null mocks (i.e., with no systematics) and finally, "known/contaminated" is for using a self-consistent covariance matrix, i.e., after applying known-1z-H512 mitigation on the contaminated mocks. 
 
 When we compare "null" cases with "contaminated cases", we find that the covariance after contamination tends to increase the error on the measurement, which makes sense as the contamination process introduces additional fluctuations. When we compare "Sys\_tot/contam" and "known/contam", we find that using a self-consistent covariance matrix tends to slightly decrease the error and bring $r_{\rm off}$ closer to $-0.4$. This self-consistent "known/contam" is shown as `known-1z\textdagger' in Table~\ref{tab:alphatable} and is our benchmark measurement in this paper.

Therefore, in comparison to the measurement with the default eBOSS QSO catalog, which gives 
\newline
\newline
$\alpha_\parallel = 1.036 \pm 0.053, \quad  \alpha_\perp = 1.014 \pm 0.033$
\newline
\newline
with a detection level of 5.29$\sigma$, after more accurate systematic mitigation, we derive 
\newline
\newline
$\alpha_\parallel = 1.026 \pm 0.053, \quad  \alpha_\perp = 1.014 \pm 0.034$
\newline
\newline
with a detection level of 5.28$\sigma$ using the new default catalogue based on ~\citet{rezaie2021}. As is the trend, the difference is very small and is 1.1\% in $\apar$. Given the much more thorough mitigation scheme does not change the constraints much, we conclude that the BAO measurement from \citet{Neveux20} is robust. 

\subsection{Varying the BAO fitting model}
In this section, we test if the dependence on the mitigation method is contingent on the freedom in the BAO fitting models. 
In Table \ref{tab:modeltable} we list the BAO constraints obtained from fitting slightly different models to the catalogues for the default eBOSS DR16 mitigation method and the benchmark `known-1z-H512'. The three alterations we make are: freeing the growth parameter $f$, adding an extra broadband term $a_3/k^2$ to each multipole, and removing the hexadecapole from the fits.  We find that for the  default catalog, freeing $f$ causes the largest shift in both $\alpha$s (around $0.5-1\%$ in $\alpha$). This is comparable to the largest shift introduced from the NN methods, although in this case $\alpha_\parallel$ is increased.  In terms of error, the greatest error reduction comes from adding the extra broadband term $1/k^2$ to each multipole, which reduces the error by about 13\% for both $\alpha$s. 

We perform a similar test on our benchmark NN catalog, fitted with a self-consistent covariance matrix.  Adding the extra $1/k^2$ term causes the largest shift in $\alpha_\parallel$ and freeing $f$ causes the largest shift in $\alpha_\perp$.  As with the default fits, adding the extra $1/k^2$ term reduces the error the most: 18\% for $\apar$ and 16\% for $\aperp$. The reduced $\chi^2$ is slightly better than the default case and the correlation between $\aperp$ and $\apar$ is still reasonably close to $-0.4$. Again, the best fit posterior $\alpha_{\rm iso}$ is more stable against all changes. 

To summarize, we find that introducing one more degree of freedom per multipole for the broadband marginalization appears to improve the BAO fit. On the other hand, we do not see that the performance of our benchmark mitigation method is sensitive to an extra degree of freedom in the BAO fitting model. 

\subsection{Mock test}
In the previous sections, we have seen that the best-fit BAO scales change only at the level of $0.2\sigma$ by applying the NN method, which we argue is statistically insignificant despite little sample variance expected between the two cases. In this section, we conduct an equivalent analysis on the EZmock catalogues in order to, first, explicitly test the significance of the offset we observed from the data, and second, to predict if the method of mitigation, i.e., the residual observational systematics, would make more difference if the statistical precision of the data were much higher. Another facet of this test is to check if the NN method, which was shown to be much more efficient for recovering an uncontaminated clustering at very low $k$ \citep{Rezaie19,rezaie2021}, does not in fact degrade the BAO feature and therefore we can use the catalogue with the NN weight for measuring both primordial non-Gaussianity and the BAO feature self-consistently. 

For the test, each set of 1000 mocks is fitted with a self-consistent covariance derived from that set. Since EZmocks are not full N-body simulations and do not account for the accurate nonlinear structure growth, we are only interested in the difference in the best fits between different scenarios. As a caveat, in generating the contaminated mocks, the contamination was applied deterministically using the default linear regression best fit to the observed imaging systematics. The default weight w$_{\rm systot}$ for the EZ mocks is also derived using the exactly the same form. That is, w$_{\rm systot}$ is a weight we could derive if we already understand the true contamination and therefore we expect this model to produce a result closest to the truth by construction. 

\begin{figure}
    \centering
    \includegraphics[width=\columnwidth]{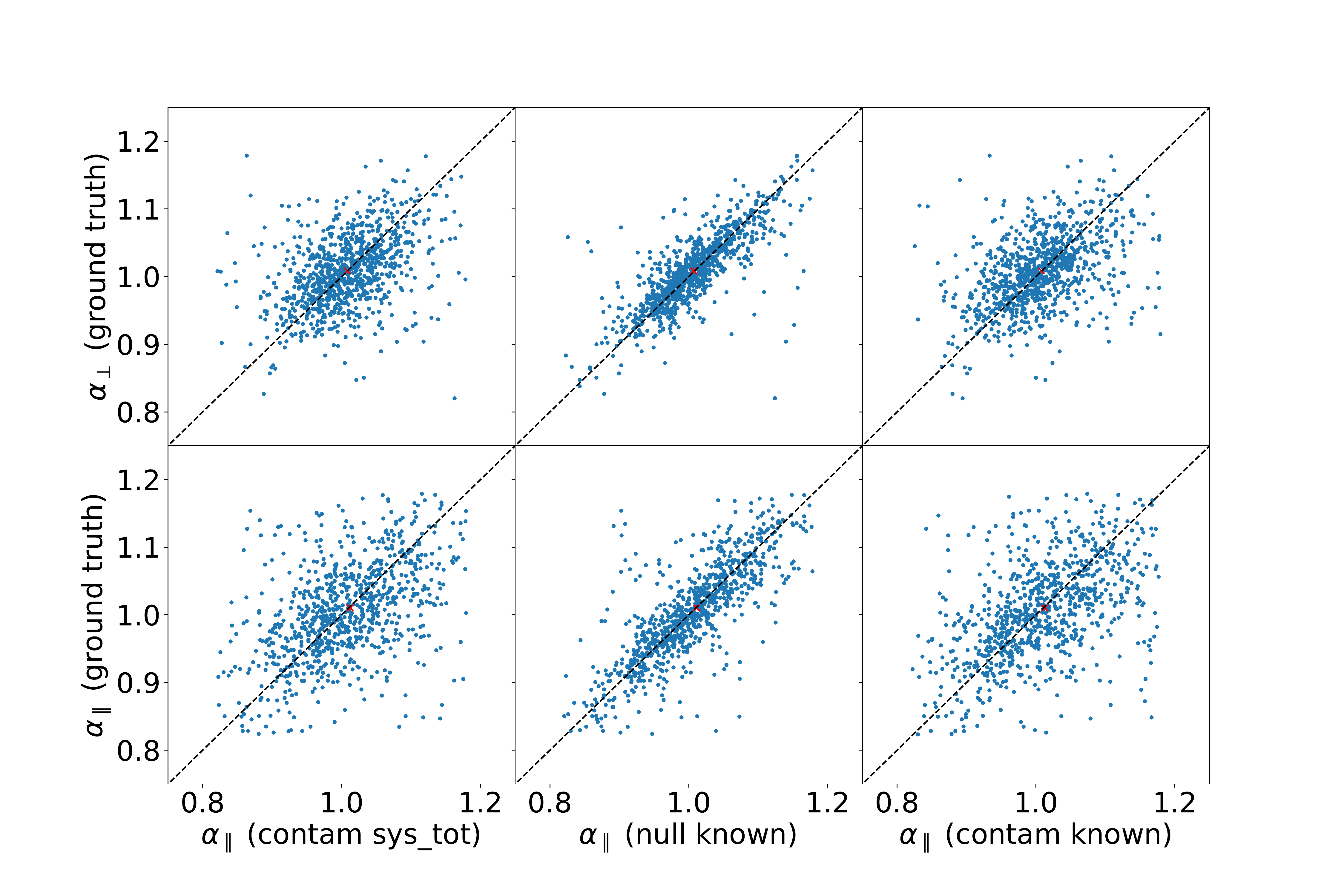}
    \caption{Spread in the best-fit values of EZmocks when applying different weights to the mocks. A covariance is derived for each case from the mocks. The y axis compares to the fits on the default EZmocks (with default covariance).  The left panels show fits done on null mocks with the NN weight (known-1z-H512).  The middle panel shows fits one contaminated mocks with the NN weight and the right panel shows fits on contaminated mocks with the default weight. Red points indicate the mean of the mocks. Mocks close to the prior boundary ($0.82<\alpha_\perp,\alpha_\parallel<1.18$) are removed.  
    }
    \label{fig:alpha_cov_spread}

\end{figure}

\begin{figure}
    \centering
    \includegraphics[width=\columnwidth]{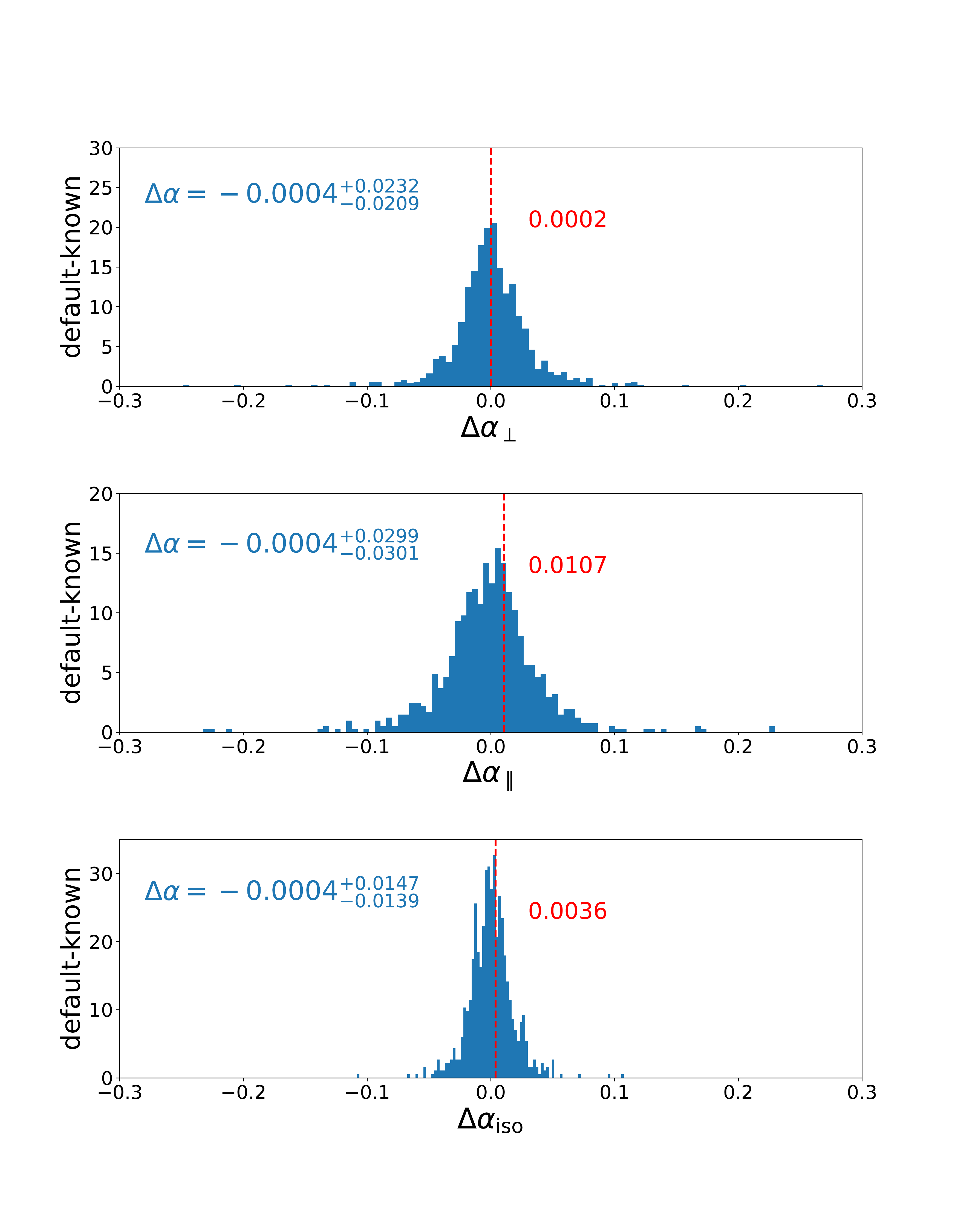}
    \caption{The distribution of alpha(default mocks) - alpha(NN mocks). The blue values indicate the median and the upper and lower error derived from the excluding the outer 32\% of the distribution.  Fits from the corresponding data catalogues are shown in red.}
    \label{fig:alpha_dist_hist}
\end{figure}

Figure \ref{fig:alpha_cov_spread} shows the scatter diagrams of the best fits when applying different mitigation, in comparison to the best fits using the ground truth power spectrum (i.e., no systematic effects and no mitigation). The plot shows that the scatter introduced in the process of contamination (i.e., the left column and the right column) is much more than the scatter introduced in the process of mitigation (the middle column), despite that the contamination is applied deterministically without adding statistical noise. The scatter plot shows many outliers near $\alpha$ $\sim$ 1.2 and 0.8, which are our prior limits; we consider these outliers have failed to detect the correct BAO feature and only retain fits within $0.82<\alpha_\perp,\alpha_\parallel<1.18$. Even after removing these, the width of the distribution is largely affected by the remaining outliers. Therefore, when we calculate the difference in $\alpha$s, we quote a median of the difference instead of an average and quote the range of 68\% of $\Delta \alpha$ around the median in Table~\ref{tab:alphashift} and Figure \ref{fig:alpha_dist_hist}.

In Figure \ref{fig:alpha_dist_hist} we look at the distribution of the difference in the best fit $\alpha$s derived from the default mocks and from contaminated mocks mitigated with the NN weights. The distribution is tighter than the statistical error of the individual mock, because the sample variance cancels out to some extent. The width and the mean of this distribution depends on how conservatively we remove the outliers.
Nevertheless, it is obvious that the offset we observed in the data between the default and known-1z-H512 (the red dashed vertical lines) falls well within the typical distribution of the mocks. Therefore, assuming this mock represents the systematic effects in the data, the offset in the best fit $\alpha$s we observed in the data (Table \ref{tab:alphatable}) is statistically insignificant. As a caveat, \citet{rezaie2021} demonstrates that the data appears to have substantially more nonlinear systematics contribution than simulated in the simulations.

The second question is if the difference due to mitigation would become significant if the statistical precision of the data improves in the future surveys. Table~\ref{tab:alphashift} shows that, after mitigation, the bias on the BAO scale (i.e., difference between the null and the known-contaminated) is $\sim$ $0.15\%$ for $\aperp$ and $\sim$ $0.2\%$ for $\apar$. The comparison between `known-null', i.e., the null catalogue mitigated with the NN method, and the ground truth shows that, if there was no systematic in the catalog, attempting to correct for the systematics may introduce a bias of 0.15\% in $\alpha_\perp$ perhaps due to overfitting (see \citet{rezaie2021} for more discussion of the overfitting issue). The legends of Figure \ref{fig:alpha_dist_hist} shows the median and the $1\sigma$ range. The median values are well within the error of the median (i.e., the $1\sigma$ range divided by the number of mocks). Therefore we expect no mitigation-dependent bias on the BAO scale at the level of 0.1\%, again assuming the property and the level of the systematics embedded in the eBOSS EZ mocks.

To summarize, assuming that the contamination introduced in the EZ mocks closely represent the systematics of the eBOSS DR16 QSO sample, we conclude that the $0.2\sigma$ of the BAO scale shift due to the NN mitigation  we detected from the data is insignificant. Also, the result implies that a bias on the BAO scale is within  $~0.1-0.2\%$ and we can use a unified catalogue for both the measurement of the primordial non-Gaussianity and the BAO scale. 
We emphasize that this does not imply that the same minimal level of systematics will influence the BAO scales in future surveys. One should repeat the test of observational systematics on the BAO scale given the survey-dependent target selection and systematics.
\begin{comment}
\begin{table}
\centering
\caption{Median shifts and the error on the median (i.e., 68\% ranges around the median in the difference in $\alpha$ divided by the number of mocks) when fitting to contaminated/null and NN/sys\_tot weighted EZ mocks. The baseline reference is taken to be null mocks with without systematic mitigation, i.e ground truth mocks.}
\normalsize
\resizebox{1.0\columnwidth}{!}{% use resizebox with textwidth
\begin{tabular}{ l | c  c  c  c }

 baseline: null & <$\Delta\alpha_\parallel$> (\%) & $ <\Delta\alpha_\perp$> (\%)   & $r_{\rm off}(\alpha_\parallel)$  & $r_{\rm off}(\alpha_\perp)$ \\ 
 \hline
  sys\_tot contam &  $-0.074^{+0.227}_{-0.236} $  & $ 0.017^{+0.166}_{-0.160} $ &  0.759  & 0.647  \\[.3cm]
  known null &   $ 0.000^{+0.103}_{-0.112} $ & $ 0.148^{+0.073}_{-0.081} $ &  0.883
  &  0.871  \\[.3cm]
  known contam &  $ 0.012^{+0.227}_{-0.244}$ & $ 0.084^{+0.167}_{-0.165}$ & 0.678 &  0.770 \\[.3cm]
  noweight contam &  $ 0.010^{+0.228}_{-0.200}$ & $ -0.108 ^{+0.162}_{-0.159}$ & 0.506 & 0.653 \\[.3cm]
  \hline  

\end{tabular}% close resizebox
}
\label{tab:alphashift}
\end{table}
\end{comment}

\begin{table}
\centering
\caption{Median shifts and the error on the median (i.e., 68\% ranges around the median in the difference in $\alpha$ divided by the number of mocks) when fitting to contaminated/null and NN/sys\_tot weighted EZ mocks. The baseline reference is taken to be null mocks with without systematic mitigation, i.e ground truth mocks.}
\normalsize
\resizebox{1.0\columnwidth}{!}{% use resizebox with textwidth
\begin{tabular}{ l | c  c  c  c }

 baseline: null & $<\Delta\alpha_\parallel>$ (\%) & $ <\Delta\alpha_\perp>$ (\%)   & $r_{\rm off}(\alpha_\parallel)$  & $r_{\rm off}(\alpha_\perp)$ \\ 
 \hline
  sys\_tot contam &  $-0.150^{+0.217}_{-0.212} $  & $ 0.049^{+0.163}_{-0.153} $ &  0.759  & 0.647  \\[.3cm]
  known null &   $ -0.049^{+0.109}_{-0.108} $ & $ 0.155^{+0.073}_{-0.080} $ &  0.883
  &  0.871  \\[.3cm]
  known contam &  $ 0.070^{+0.226}_{-0.198}$ & $ 0.115^{+0.157}_{-0.157}$ & 0.678 &  0.770 \\[.3cm]
  noweight contam &  $ 0.053^{+0.217}_{-0.196}$ & $ -0.093 ^{+0.158}_{-0.151}$ & 0.506 & 0.653 \\[.3cm]
  \hline

\end{tabular}% close resizebox
}
\label{tab:alphashift}
\end{table}

\section{Conclusion}
\label{sec:conclusion}

The BAO signal is a very powerful cosmological probe in investigating the large scale structure of the universe.  As upcoming next-generation surveys will probe deeper into the cosmos, they will allow for the most precise measurements of the BAO through the galaxy clustering data. However, the observed galaxy clustering is affected by the experimental setups, such as obscuring dust in the galactic plane and stars in the foreground of distant galaxies. Previous surveys often used linear regression models to calculate systematic weights to mitigate this effect, and it is believed that the BAO measurement is in general fairly robust against observational systematics in the clustering data.  While this may be the case, new methods based on neural networks are being developed to deal with observational systematics, especially on large scales.  These methods can account for nonlinearities in the data and are in general more flexible.  It is important to confirm that new mitigation schemes will not bias the clustering on scales relevant to the BAO. Such a neural network is developed in \cite{Rezaie19} and utilized in this paper, where we employ several different setups of hyperparameters and test the effect on the BAO signal in the eBOSS DR16 QSO clustering data.  We summarize our findings here.

In general, the systematic weights derived from NN mitigation did not significantly alter the BAO shape or location compared to the linear regression method, evidenced by the fact that the $\alpha$ parameters did not differ significantly across the fits. Changes in best-fit $\alpha_\perp$ and $\alpha_\parallel$ were well within the error from the posterior, at most $0.2\sigma$. This result supports the robustness of the eBOSS BAO result reported in \citet{Neveux20} against observational systematics. 

In detail, all fits to NN weighted catalogues have a slightly larger error and $\chi^2$ and produce a smaller $\alpha_\parallel$ and $\alpha_\perp$.  The isotropic $\alpha$s appear to be more stable across different fits. There is no clear trend in the $\alpha$ cross-correlation coefficients with the varying NN setups. 

We then examine the propagation of errors through the covariance matrix used in the fitting process.  We derive a covariance matrix from 1000 EZ mocks that have been treated with our benchmark NN mitigation, and perform a fit on the data that has undergone the same mitigation.  We find that the best fit $\alpha_\parallel$ is decreased by 1.3\%, and $\alpha_\perp$ remains unchanged. From this test, we conclude that the previous BAO measurement from \cite{Neveux20} is robust.

We also tested if the dependence on the mitigation method is contingent on the freedom in the BAO fitting models. We find that introducing one more degree of freedom per multipole for the broadband marginalization appears to improve the BAO fit by around 15-18\%. On the other hand, we do not see that the performance of our benchmark mitigation method is sensitive to an extra degree of freedom in the BAO fitting model.

We conduct a mock test to quantify the significance of the difference in $\alpha$s we observed from the data. Assuming that the contamination introduced in the EZ mocks closely represent the systematics of the eBOSS DR16 QSO sample, we conclude that the $0.2\sigma$ of the BAO scale shift due to the NN mitigation we detected from the data is insignificant. Our mock result also implies that a bias on the BAO scale would be within  $~0.1-0.2\%$, i.e., negligible if we had a much more precise data and also we can use a unified catalogue for both the measurement of the primordial non-Gaussianity and the BAO scale. A similar test of observational systematics on the BAO scale  should be conducted for future surveys given the survey-dependent target selection and systematics.

Additionally, a fitting attempt was made on the high-z sample of data (see Appendix \ref{App:highz}), which covered QSOs out to redshift 3.5.  This a less dense sample, as fainter galaxies are harder to detect.  As such, the data is a lot noisier.  Given that there are no EZmocks that cover this redshift range, we rely on analytical approximations of the covariance matrix dervied from the standard sample EZmocks.  A fit was not able to produce physically meaningful values; in particular $\alpha_\perp$ was driven to small values at the edge of the prior range.  With more data and a higher signal-to-noise ratio, a better fit might be possible at higher $z$.

The era of big data in astronomy is here.  New surveys will peer into the universe as never before and produce massive amounts of data that will help cosmologists uncover the secrets of dark energy.  In order for the most accurate measurements to be made, experiments must be aware of systematic errors caused by outside factors and a thorough test of systematics such as this work should be conducted.  While the BAO is believed to be robust and this work indeed confirms it within the eBOSS precision, it will be important to carefully consider how to produce systematic weights for future projects.

\section*{Acknowledgements}
G. M. is supported by the U.S.~Department of Energy, Office of Science, Office of High Energy Physics under DE-SC0019091. M.R. and H.-J.S. are supported by the U.S.~Department of Energy, Office of Science, Office of High Energy Physics under DE-SC0014329 and DE-SC0019091. E.-M. M. acknowledges support from the European Research Council (ERC) under the European Union’s Horizon 2020 research and innovation programme (grant agreement No 693024).
 We acknowledge support and resources from the Ohio Supercomputer Center \citep[OSC;][]{OSC87}. Specifically, this work utilized more than $359000$ core hours of the Owens cluster \citep{Owens2016}. M.R. is grateful for help from Xia Wang, Antonio Marcum, and Yu Feng.  This work made substantial use of open-source software and modules, such as Pytorch, Nbodykit, \textsc{HEALPix}, Fitsio, Scikit-Learn, NumPy, SciPy, Pandas, IPython, Jupyter, and GitHub. G.R. acknowledges support from the National Research Foundation of Korea (NRF) through Grants No. 2017R1E1A1A01077508 and No.2020R1A2C1005655 funded by the Korean Ministry of Education, Science and Technology (MoEST).

Funding for the Sloan Digital Sky 
Survey IV has been provided by the 
Alfred P. Sloan Foundation, the U.S. 
Department of Energy Office of 
Science, and the Participating 
Institutions. SDSS-IV acknowledges support and 
resources from the Center for High 
Performance Computing  at the 
University of Utah. The SDSS 
website is www.sdss.org. This work also relied on resources provided to the eBOSS Collaboration by the National Energy Research Scientific Computing Center (NERSC). NERSC is a U.S. Department of Energy Office of Science User Facility operated under Contract No. DE-AC02-05CH11231.

SDSS-IV is managed by the 
Astrophysical Research Consortium 
for the Participating Institutions 
of the SDSS Collaboration including 
the Brazilian Participation Group, 
the Carnegie Institution for Science, 
Carnegie Mellon University, Center for 
Astrophysics | Harvard \& 
Smithsonian, the Chilean Participation 
Group, the French Participation Group, 
Instituto de Astrof\'isica de 
Canarias, The Johns Hopkins 
University, Kavli Institute for the 
Physics and Mathematics of the 
Universe (IPMU) / University of 
Tokyo, the Korean Participation Group, 
Lawrence Berkeley National Laboratory, 
Leibniz Institut f\"ur Astrophysik 
Potsdam (AIP),  Max-Planck-Institut 
f\"ur Astronomie (MPIA Heidelberg), 
Max-Planck-Institut f\"ur 
Astrophysik (MPA Garching), 
Max-Planck-Institut f\"ur 
Extraterrestrische Physik (MPE), 
National Astronomical Observatories of 
China, New Mexico State University, 
New York University, University of 
Notre Dame, Observat\'ario 
Nacional / MCTI, The Ohio State 
University, Pennsylvania State 
University, Shanghai 
Astronomical Observatory, United 
Kingdom Participation Group, 
Universidad Nacional Aut\'onoma 
de M\'exico, University of Arizona, 
University of Colorado Boulder, 
University of Oxford, University of 
Portsmouth, University of Utah, 
University of Virginia, University 
of Washington, University of 
Wisconsin, Vanderbilt University, 
and Yale University.
 
%%%%%%%%%%%%%%%%%%%%%%%%%%%%%%%%%%%%%%%%%%%%%%%%%%


\bibliographystyle{mnras}
\bibliography{biblio}

\begin{thebibliography}{}
\makeatletter
\relax
\def\mn@urlcharsother{\let\do\@makeother \do\$\do\&\do\#\do\^\do\_\do\%\do\~}
\def\mn@doi{\begingroup\mn@urlcharsother \@ifnextchar [ {\mn@doi@}
  {\mn@doi@[]}}
\def\mn@doi@[#1]#2{\def\@tempa{#1}\ifx\@tempa\@empty \href
  {http://dx.doi.org/#2} {doi:#2}\else \href {http://dx.doi.org/#2} {#1}\fi
  \endgroup}
\def\mn@eprint#1#2{\mn@eprint@#1:#2::\@nil}
\def\mn@eprint@arXiv#1{\href {http://arxiv.org/abs/#1} {{\tt arXiv:#1}}}
\def\mn@eprint@dblp#1{\href {http://dblp.uni-trier.de/rec/bibtex/#1.xml}
  {dblp:#1}}
\def\mn@eprint@#1:#2:#3:#4\@nil{\def\@tempa {#1}\def\@tempb {#2}\def\@tempc
  {#3}\ifx \@tempc \@empty \let \@tempc \@tempb \let \@tempb \@tempa \fi \ifx
  \@tempb \@empty \def\@tempb {arXiv}\fi \@ifundefined
  {mn@eprint@\@tempb}{\@tempb:\@tempc}{\expandafter \expandafter \csname
  mn@eprint@\@tempb\endcsname \expandafter{\@tempc}}}

\bibitem[\protect\citeauthoryear{{Ahumada} et~al.,}{{Ahumada}
  et~al.}{2020}]{Ahumada2020ApJS..249....3A}
{Ahumada} R.,  et~al., 2020, \mn@doi [\apjs] {10.3847/1538-4365/ab929e}, \href
  {https://ui.adsabs.harvard.edu/abs/2020ApJS..249....3A} {249, 3}

\bibitem[\protect\citeauthoryear{{Alam} et~al.,}{{Alam} et~al.}{2020}]{Alam20}
{Alam} S.,  et~al., 2020, arXiv e-prints, \href
  {https://ui.adsabs.harvard.edu/abs/2020arXiv200709004A} {p. arXiv:2007.09004}

\bibitem[\protect\citeauthoryear{{Alcock} \& {Paczynski}}{{Alcock} \&
  {Paczynski}}{1979}]{AP79}
{Alcock} C.,  {Paczynski} B.,  1979, \mn@doi [\nat] {10.1038/281358a0}, \href
  {https://ui.adsabs.harvard.edu/abs/1979Natur.281..358A} {281, 358}

\bibitem[\protect\citeauthoryear{{Anderson}, {Aubourg}, {Bailey}
  et~al.}{{Anderson} et~al.}{2014}]{Anderson14}
{Anderson} L.,  {Aubourg} E.,  {Bailey} S.,   et~al., 2014, \mn@doi [Monthly
  Notices of the Royal Astronomical Society] {10.1093/mnras/stt2206}, \href
  {https://ui.adsabs.harvard.edu/abs/2014MNRAS.439...83A} {439, 83}

\bibitem[\protect\citeauthoryear{{Ata} et~al.,}{{Ata} et~al.}{2018}]{Ata17}
{Ata} M.,  et~al., 2018, \mn@doi [\mnras] {10.1093/mnras/stx2630}, \href
  {https://ui.adsabs.harvard.edu/abs/2018MNRAS.473.4773A} {473, 4773}

\bibitem[\protect\citeauthoryear{{Ballinger}, {Peacock}  \&
  {Heavens}}{{Ballinger} et~al.}{1996}]{BPH96}
{Ballinger} W.~E.,  {Peacock} J.~A.,   {Heavens} A.~F.,  1996, \mn@doi [\mnras]
  {10.1093/mnras/282.3.877}, \href
  {https://ui.adsabs.harvard.edu/abs/1996MNRAS.282..877B} {282, 877}

\bibitem[\protect\citeauthoryear{Bautista et~al.,}{Bautista
  et~al.}{2018}]{bautista2018sdss}
Bautista J.~E.,  et~al., 2018, The Astrophysical Journal, 863, 110

\bibitem[\protect\citeauthoryear{{Beutler}, {Blake}, {Koda}, {Mar{\'\i}n},
  {Seo}, {Cuesta}  \& {Schneider}}{{Beutler} et~al.}{2016}]{Beutler16}
{Beutler} F.,  {Blake} C.,  {Koda} J.,  {Mar{\'\i}n} F.~A.,  {Seo} H.-J.,
  {Cuesta} A.~J.,   {Schneider} D.~P.,  2016, \mn@doi [Monthly Notices of the
  Royal Astronomical Society] {10.1093/mnras/stv1943}, \href
  {https://ui.adsabs.harvard.edu/abs/2016MNRAS.455.3230B} {455, 3230}

\bibitem[\protect\citeauthoryear{{Beutler}, {Castorina}  \& {Zhang}}{{Beutler}
  et~al.}{2019}]{Beutler19}
{Beutler} F.,  {Castorina} E.,   {Zhang} P.,  2019, \mn@doi [\jcap]
  {10.1088/1475-7516/2019/03/040}, \href
  {https://ui.adsabs.harvard.edu/abs/2019JCAP...03..040B} {2019, 040}

\bibitem[\protect\citeauthoryear{{Beutler et al. in preparation}}{{Beutler et
  al. in preparation}}{2021}]{Beutler2021}
{Beutler et al. in preparation} 2021, \mn@doi [\mnras] {XX}, \href {XXXXX} {XX,
  XX}

\bibitem[\protect\citeauthoryear{{Bianchi}, {Gil-Mar{\'\i}n}, {Ruggeri}  \&
  {Percival}}{{Bianchi} et~al.}{2015}]{Bianchi15}
{Bianchi} D.,  {Gil-Mar{\'\i}n} H.,  {Ruggeri} R.,   {Percival} W.~J.,  2015,
  \mn@doi [\mnras] {10.1093/mnrasl/slv090}, \href
  {https://ui.adsabs.harvard.edu/abs/2015MNRAS.453L..11B} {453, L11}

\bibitem[\protect\citeauthoryear{{Blanton} et~al.,}{{Blanton}
  et~al.}{2017}]{SDSS4}
{Blanton} M.~R.,  et~al., 2017, \mn@doi [\aj] {10.3847/1538-3881/aa7567}, \href
  {https://ui.adsabs.harvard.edu/abs/2017AJ....154...28B} {154, 28}

\bibitem[\protect\citeauthoryear{{Chuang}, {Kitaura}, {Prada}, {Zhao}  \&
  {Yepes}}{{Chuang} et~al.}{2015}]{Chuang15}
{Chuang} C.-H.,  {Kitaura} F.-S.,  {Prada} F.,  {Zhao} C.,   {Yepes} G.,  2015,
  \mn@doi [Monthly Notices of the Royal Astronomical Society]
  {10.1093/mnras/stu2301}, \href
  {https://ui.adsabs.harvard.edu/abs/2015MNRAS.446.2621C} {446, 2621}

\bibitem[\protect\citeauthoryear{{DESI Collaboration} et~al.,}{{DESI
  Collaboration} et~al.}{2016}]{DESI16}
{DESI Collaboration} et~al., 2016, arXiv e-prints, \href
  {https://ui.adsabs.harvard.edu/abs/2016arXiv161100036D} {p. arXiv:1611.00036}

\bibitem[\protect\citeauthoryear{{Dawson} et~al.,}{{Dawson}
  et~al.}{2013}]{BOSS13}
{Dawson} K.~S.,  et~al., 2013, \mn@doi [\aj] {10.1088/0004-6256/145/1/10},
  \href {https://ui.adsabs.harvard.edu/abs/2013AJ....145...10D} {145, 10}

\bibitem[\protect\citeauthoryear{{Dawson} et~al.,}{{Dawson}
  et~al.}{2016}]{eBOSS16}
{Dawson} K.~S.,  et~al., 2016, \mn@doi [\aj] {10.3847/0004-6256/151/2/44},
  \href {https://ui.adsabs.harvard.edu/abs/2016AJ....151...44D} {151, 44}

\bibitem[\protect\citeauthoryear{{Eisenstein} \& {Hu}}{{Eisenstein} \&
  {Hu}}{1998}]{Eisenstein98}
{Eisenstein} D.~J.,  {Hu} W.,  1998, \mn@doi [The Astrophysical Journal]
  {10.1086/305424}, \href
  {https://ui.adsabs.harvard.edu/abs/1998ApJ...496..605E} {496, 605}

\bibitem[\protect\citeauthoryear{{Eisenstein}, {Zehavi}, {Hogg}
  et~al.}{{Eisenstein} et~al.}{2005}]{Eisenstein05}
{Eisenstein} D.~J.,  {Zehavi} I.,  {Hogg} D.~W.,   et~al., 2005, \mn@doi [The
  Astrophysical Journal] {10.1086/466512}, \href
  {https://ui.adsabs.harvard.edu/abs/2005ApJ...633..560E} {633, 560}

\bibitem[\protect\citeauthoryear{Eisenstein et~al.,}{Eisenstein
  et~al.}{2011}]{eisenstein2011sdss}
Eisenstein D.~J.,  et~al., 2011, The Astronomical Journal, 142, 72

\bibitem[\protect\citeauthoryear{{Feldman}, {Kaiser}  \& {Peacock}}{{Feldman}
  et~al.}{1994}]{FKP94}
{Feldman} H.~A.,  {Kaiser} N.,   {Peacock} J.~A.,  1994, \mn@doi [Astrophysical
  Journal] {10.1086/174036}, \href
  {https://ui.adsabs.harvard.edu/abs/1994ApJ...426...23F} {426, 23}

\bibitem[\protect\citeauthoryear{{Font-Ribera}, {McDonald}, {Mostek}, {Reid},
  {Seo}  \& {Slosar}}{{Font-Ribera} et~al.}{2014}]{DESIFisher}
{Font-Ribera} A.,  {McDonald} P.,  {Mostek} N.,  {Reid} B.~A.,  {Seo} H.-J.,
  {Slosar} A.,  2014, \mn@doi [\jcap] {10.1088/1475-7516/2014/05/023}, \href
  {https://ui.adsabs.harvard.edu/abs/2014JCAP...05..023F} {2014, 023}

\bibitem[\protect\citeauthoryear{{Foreman-Mackey}, {Hogg}, {Lang}  \&
  {Goodman}}{{Foreman-Mackey} et~al.}{2013}]{Emcee13}
{Foreman-Mackey} D.,  {Hogg} D.~W.,  {Lang} D.,   {Goodman} J.,  2013, \mn@doi
  [Publications of the ASP] {10.1086/670067}, \href
  {https://ui.adsabs.harvard.edu/abs/2013PASP..125..306F} {125, 306}

\bibitem[\protect\citeauthoryear{Fukugita, Shimasaku, Ichikawa, Gunn
  et~al.}{Fukugita et~al.}{1996}]{fukugita1996sloan}
Fukugita M.,  Shimasaku K.,  Ichikawa T.,  Gunn J.,   et~al., 1996, Technical
  report, The Sloan digital sky survey photometric system.
SCAN-9601313

\bibitem[\protect\citeauthoryear{{Gaia Collaboration} et~al.,}{{Gaia
  Collaboration} et~al.}{2018}]{gaia2018}
{Gaia Collaboration} et~al., 2018, \mn@doi [AAP] {10.1051/0004-6361/201833051},
  \href {https://ui.adsabs.harvard.edu/abs/2018A&A...616A...1G} {616, A1}

\bibitem[\protect\citeauthoryear{{Gunn} et~al.,}{{Gunn}
  et~al.}{2006}]{SloanTelescope}
{Gunn} J.~E.,  et~al., 2006, \mn@doi [\aj] {10.1086/500975}, \href
  {https://ui.adsabs.harvard.edu/abs/2006AJ....131.2332G} {131, 2332}

\bibitem[\protect\citeauthoryear{{HI4PI Collaboration} et~al.,}{{HI4PI
  Collaboration} et~al.}{2016}]{Hi4pi2016}
{HI4PI Collaboration} et~al., 2016, \mn@doi [AAP]
  {10.1051/0004-6361/201629178}, \href
  {https://ui.adsabs.harvard.edu/abs/2016A&A...594A.116H} {594, A116}

\bibitem[\protect\citeauthoryear{{Hand}, {Li}, {Slepian}  \& {Seljak}}{{Hand}
  et~al.}{2017}]{Hand17}
{Hand} N.,  {Li} Y.,  {Slepian} Z.,   {Seljak} U.,  2017, \mn@doi [Journal of
  Cosmology and Astroparticle Physics] {10.1088/1475-7516/2017/07/002}, \href
  {https://ui.adsabs.harvard.edu/abs/2017JCAP...07..002H} {2017, 002}

\bibitem[\protect\citeauthoryear{{Hand}, {Feng}, {Beutler}, {Li}, {Modi},
  {Seljak}  \& {Slepian}}{{Hand} et~al.}{2018}]{Hand18}
{Hand} N.,  {Feng} Y.,  {Beutler} F.,  {Li} Y.,  {Modi} C.,  {Seljak} U.,
  {Slepian} Z.,  2018, \mn@doi [Astronomical Journal]
  {10.3847/1538-3881/aadae0}, \href
  {https://ui.adsabs.harvard.edu/abs/2018AJ....156..160H} {156, 160}

\bibitem[\protect\citeauthoryear{{Hartlap}, {Simon}  \& {Schneider}}{{Hartlap}
  et~al.}{2007}]{Hartlap07}
{Hartlap} J.,  {Simon} P.,   {Schneider} P.,  2007, \mn@doi [\aap]
  {10.1051/0004-6361:20066170}, \href
  {https://ui.adsabs.harvard.edu/abs/2007A&A...464..399H} {464, 399}

\bibitem[\protect\citeauthoryear{Kalus, Percival, Bacon, Mueller, Samushia,
  Verde, Ross  \& Bernal}{Kalus et~al.}{2019}]{kalus2019map}
Kalus B.,  Percival W.,  Bacon D.,  Mueller E.,  Samushia L.,  Verde L.,  Ross
  A.,   Bernal J.,  2019, Monthly Notices of the Royal Astronomical Society,
  482, 453

\bibitem[\protect\citeauthoryear{{Lyke} et~al.,}{{Lyke}
  et~al.}{2020}]{lyke2020ApJS..250....8L}
{Lyke} B.~W.,  et~al., 2020, \mn@doi [\apjs] {10.3847/1538-4365/aba623}, \href
  {https://ui.adsabs.harvard.edu/abs/2020ApJS..250....8L} {250, 8}

\bibitem[\protect\citeauthoryear{{Mueller et al.}}{{Mueller et
  al.}}{2021}]{mueller2021}
{Mueller et al.} E.,  2021, \mn@doi [\mnras] {XX}, \href {XXXXX} {XX, XX}

\bibitem[\protect\citeauthoryear{{Myers} et~al.,}{{Myers}
  et~al.}{2015}]{myers2015ApJS..221...27M}
{Myers} A.~D.,  et~al., 2015, \mn@doi [\apjs] {10.1088/0067-0049/221/2/27},
  \href {https://ui.adsabs.harvard.edu/abs/2015ApJS..221...27M} {221, 27}

\bibitem[\protect\citeauthoryear{{Neveux}, {Burtin}  \& et al.}{{Neveux}
  et~al.}{2020}]{Neveux20}
{Neveux} R.,  {Burtin} E.,   et al. 2020, Manuscript in Preparation

\bibitem[\protect\citeauthoryear{{Ohio Supercomputer Center}}{{Ohio
  Supercomputer Center}}{1987}]{OSC87}
{Ohio Supercomputer Center} 1987, {Ohio Supercomputer Center}, \url
  {http://osc.edu/ark:/19495/f5s1ph73}

\bibitem[\protect\citeauthoryear{{Ohio Supercomputer Center}}{{Ohio
  Supercomputer Center}}{2016}]{Owens2016}
{Ohio Supercomputer Center} 2016, Owens Supercomputer, \url
  {http://osc.edu/ark:/19495/hpc6h5b1}

\bibitem[\protect\citeauthoryear{{Percival} et~al.,}{{Percival}
  et~al.}{2014}]{Percival14}
{Percival} W.~J.,  et~al., 2014, \mn@doi [\mnras] {10.1093/mnras/stu112}, \href
  {https://ui.adsabs.harvard.edu/abs/2014MNRAS.439.2531P} {439, 2531}

\bibitem[\protect\citeauthoryear{{Pullen} \& {Hirata}}{{Pullen} \&
  {Hirata}}{2013}]{Pullen13}
{Pullen} A.~R.,  {Hirata} C.~M.,  2013, \mn@doi [\pasp] {10.1086/671189}, \href
  {https://ui.adsabs.harvard.edu/abs/2013PASP..125..705P} {125, 705}

\bibitem[\protect\citeauthoryear{{Rezaie}, {Seo}, {Ross}  \&
  {Bunescu}}{{Rezaie} et~al.}{2019}]{Rezaie19}
{Rezaie} M.,  {Seo} H.-J.,  {Ross} A.~J.,   {Bunescu} R.~C.,  2019, arXiv
  e-prints, \href {https://ui.adsabs.harvard.edu/abs/2019arXiv190711355R} {p.
  arXiv:1907.11355}

\bibitem[\protect\citeauthoryear{{Rezaie et al.}}{{Rezaie et
  al.}}{2021}]{rezaie2021}
{Rezaie et al.} E.,  2021, \mn@doi [\mnras] {XX}, \href {XXXXX} {XX, XX}

\bibitem[\protect\citeauthoryear{{Ross} et~al.,}{{Ross} et~al.}{2013}]{Ross13}
{Ross} A.~J.,  et~al., 2013, \mn@doi [\mnras] {10.1093/mnras/sts094}, \href
  {https://ui.adsabs.harvard.edu/abs/2013MNRAS.428.1116R} {428, 1116}

\bibitem[\protect\citeauthoryear{{Ross} et~al.,}{{Ross}
  et~al.}{2017}]{Ross2017}
{Ross} A.~J.,  et~al., 2017, \mn@doi [\mnras] {10.1093/mnras/stw2372}, \href
  {https://ui.adsabs.harvard.edu/abs/2017MNRAS.464.1168R} {464, 1168}

\bibitem[\protect\citeauthoryear{{Ross} et~al.,}{{Ross}
  et~al.}{2020}]{Ross2020}
{Ross} A.~J.,  et~al., 2020, \mn@doi [\mnras] {10.1093/mnras/staa2416}, \href
  {https://ui.adsabs.harvard.edu/abs/2020MNRAS.498.2354R} {498, 2354}

\bibitem[\protect\citeauthoryear{{Rossi} et~al.,}{{Rossi}
  et~al.}{2020}]{Rossi20}
{Rossi} G.,  et~al., 2020, \mn@doi [\mnras] {10.1093/mnras/staa3955}, \href
  {https://ui.adsabs.harvard.edu/abs/2020MNRAS.tmp.3762R} {}

\bibitem[\protect\citeauthoryear{Schlegel, Finkbeiner  \& Davis}{Schlegel
  et~al.}{1998}]{schlegel1998maps}
Schlegel D.~J.,  Finkbeiner D.~P.,   Davis M.,  1998, The Astrophysical
  Journal, 500, 525

\bibitem[\protect\citeauthoryear{{Sefusatti}, {Crocce}, {Scoccimarro}  \&
  {Couchman}}{{Sefusatti} et~al.}{2016}]{Interlacing}
{Sefusatti} E.,  {Crocce} M.,  {Scoccimarro} R.,   {Couchman} H.~M.~P.,  2016,
  \mn@doi [\mnras] {10.1093/mnras/stw1229}, \href
  {https://ui.adsabs.harvard.edu/abs/2016MNRAS.460.3624S} {460, 3624}

\bibitem[\protect\citeauthoryear{{Seo} \& {Eisenstein}}{{Seo} \&
  {Eisenstein}}{2007}]{SeoFisher2007}
{Seo} H.-J.,  {Eisenstein} D.~J.,  2007, \mn@doi [\apj] {10.1086/519549}, \href
  {https://ui.adsabs.harvard.edu/abs/2007ApJ...665...14S} {665, 14}

\bibitem[\protect\citeauthoryear{Smee et~al.,}{Smee
  et~al.}{2013}]{smee2013multi}
Smee S.~A.,  et~al., 2013, The Astronomical Journal, 146, 32

\bibitem[\protect\citeauthoryear{{Smith} et~al.,}{{Smith}
  et~al.}{2020}]{Smith20}
{Smith} A.,  et~al., 2020, \mn@doi [\mnras] {10.1093/mnras/staa2825}, \href
  {https://ui.adsabs.harvard.edu/abs/2020MNRAS.499..269S} {499, 269}

\bibitem[\protect\citeauthoryear{{Thomas}, {Abdalla}  \& {Lahav}}{{Thomas}
  et~al.}{2011}]{Thomas2011PhRvL.106x1301T}
{Thomas} S.~A.,  {Abdalla} F.~B.,   {Lahav} O.,  2011, \mn@doi [\prl]
  {10.1103/PhysRevLett.106.241301}, \href
  {https://ui.adsabs.harvard.edu/abs/2011PhRvL.106x1301T} {106, 241301}

\bibitem[\protect\citeauthoryear{{Wang}, {Beutler}  \& {Bacon}}{{Wang}
  et~al.}{2020}]{Wang2020MNRAS.499.2598W}
{Wang} M.~S.,  {Beutler} F.,   {Bacon} D.,  2020, \mn@doi [\mnras]
  {10.1093/mnras/staa2998}, \href
  {https://ui.adsabs.harvard.edu/abs/2020MNRAS.499.2598W} {499, 2598}

\bibitem[\protect\citeauthoryear{{Weinberg}, {Mortonson}, {Eisenstein},
  {Hirata}, {Riess}  \& {Rozo}}{{Weinberg} et~al.}{2013}]{Weinberg13}
{Weinberg} D.~H.,  {Mortonson} M.~J.,  {Eisenstein} D.~J.,  {Hirata} C.,
  {Riess} A.~G.,   {Rozo} E.,  2013, \mn@doi [Physics Reports]
  {10.1016/j.physrep.2013.05.001}, \href
  {https://ui.adsabs.harvard.edu/abs/2013PhR...530...87W} {530, 87}

\bibitem[\protect\citeauthoryear{Wright et~al.,}{Wright
  et~al.}{2010}]{wright2010wide}
Wright E.~L.,  et~al., 2010, The Astronomical Journal, 140, 1868

\bibitem[\protect\citeauthoryear{{Yamamoto}, {Nakamichi}, {Kamino}, {Bassett}
  \& {Nishioka}}{{Yamamoto} et~al.}{2006}]{Yamamoto06}
{Yamamoto} K.,  {Nakamichi} M.,  {Kamino} A.,  {Bassett} B.~A.,   {Nishioka}
  H.,  2006, \mn@doi [\pasj] {10.1093/pasj/58.1.93}, \href
  {https://ui.adsabs.harvard.edu/abs/2006PASJ...58...93Y} {58, 93}

\bibitem[\protect\citeauthoryear{York et~al.,}{York
  et~al.}{2000}]{york2000sloan}
York D.~G.,  et~al., 2000, The Astronomical Journal, 120, 1579

\bibitem[\protect\citeauthoryear{{Zel'Dovich}}{{Zel'Dovich}}{1970}]{Zel70}
{Zel'Dovich} Y.~B.,  1970, \aap, \href
  {https://ui.adsabs.harvard.edu/abs/1970A&A.....5...84Z} {500, 13}

\bibitem[\protect\citeauthoryear{{Zhao} et~al.,}{{Zhao}
  et~al.}{2021}]{EZmock20}
{Zhao} C.,  et~al., 2021, \mn@doi [\mnras] {10.1093/mnras/stab510}, \href
  {https://ui.adsabs.harvard.edu/abs/2021MNRAS.503.1149Z} {503, 1149}

\bibitem[\protect\citeauthoryear{{eBOSS Collaboration} et~al.,}{{eBOSS
  Collaboration} et~al.}{2020}]{eBOSS2020}
{eBOSS Collaboration} et~al., 2020, arXiv e-prints, \href
  {https://ui.adsabs.harvard.edu/abs/2020arXiv200708991E} {p. arXiv:2007.08991}

\makeatother
\end{thebibliography}

%%%%%%%%%%%%%%%%%%%%%%%%%%%%%%%%%%%%%%%%%%%%%%%%%%

%%%%%%%%%%%%%%%%% APPENDICES %%%%%%%%%%%%%%%%%%%%%

\appendix

\section{H256 Power Spectra}
\label{App:256}

\begin{figure}
    \centering
    \includegraphics[width=\columnwidth]{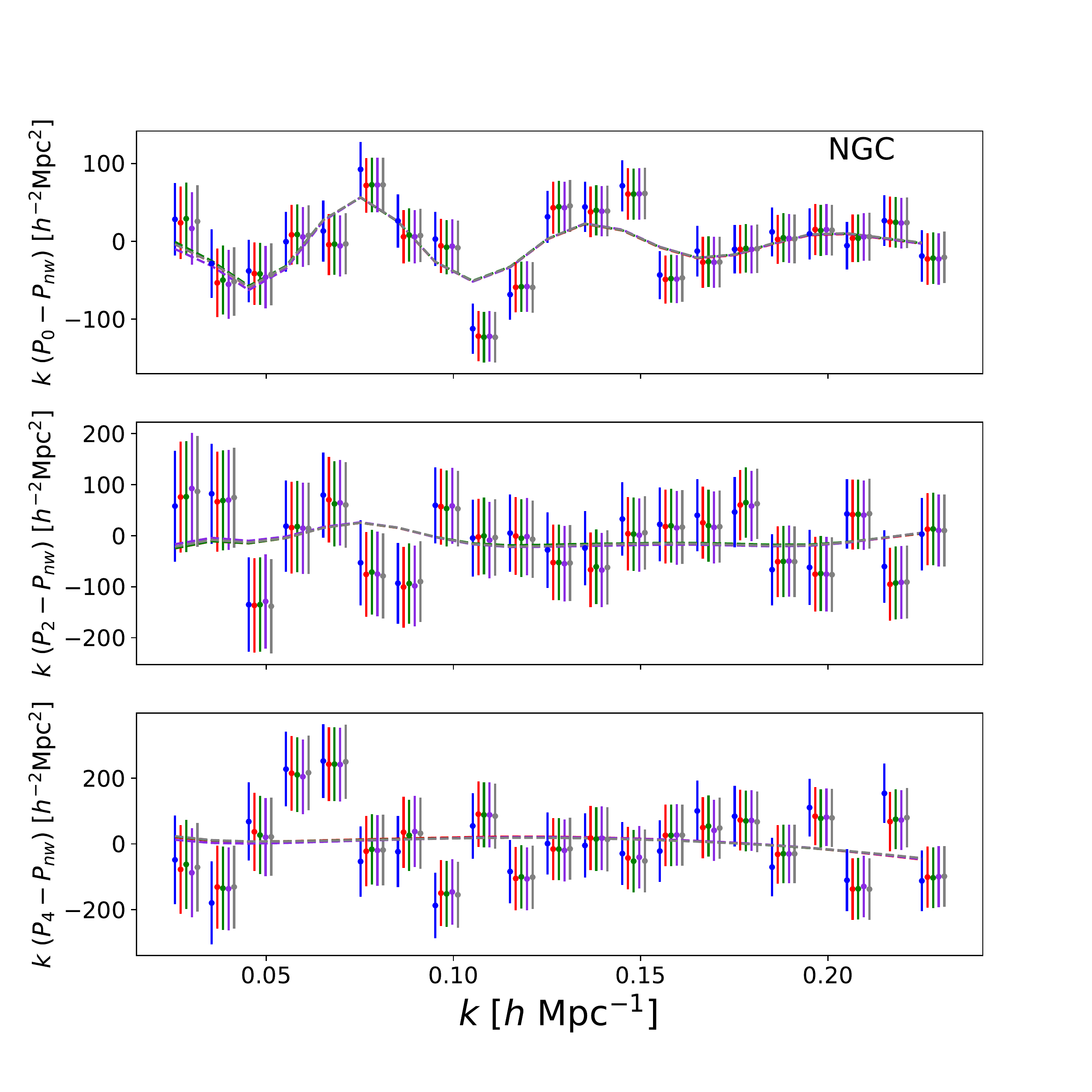}
    \includegraphics[width=\columnwidth]{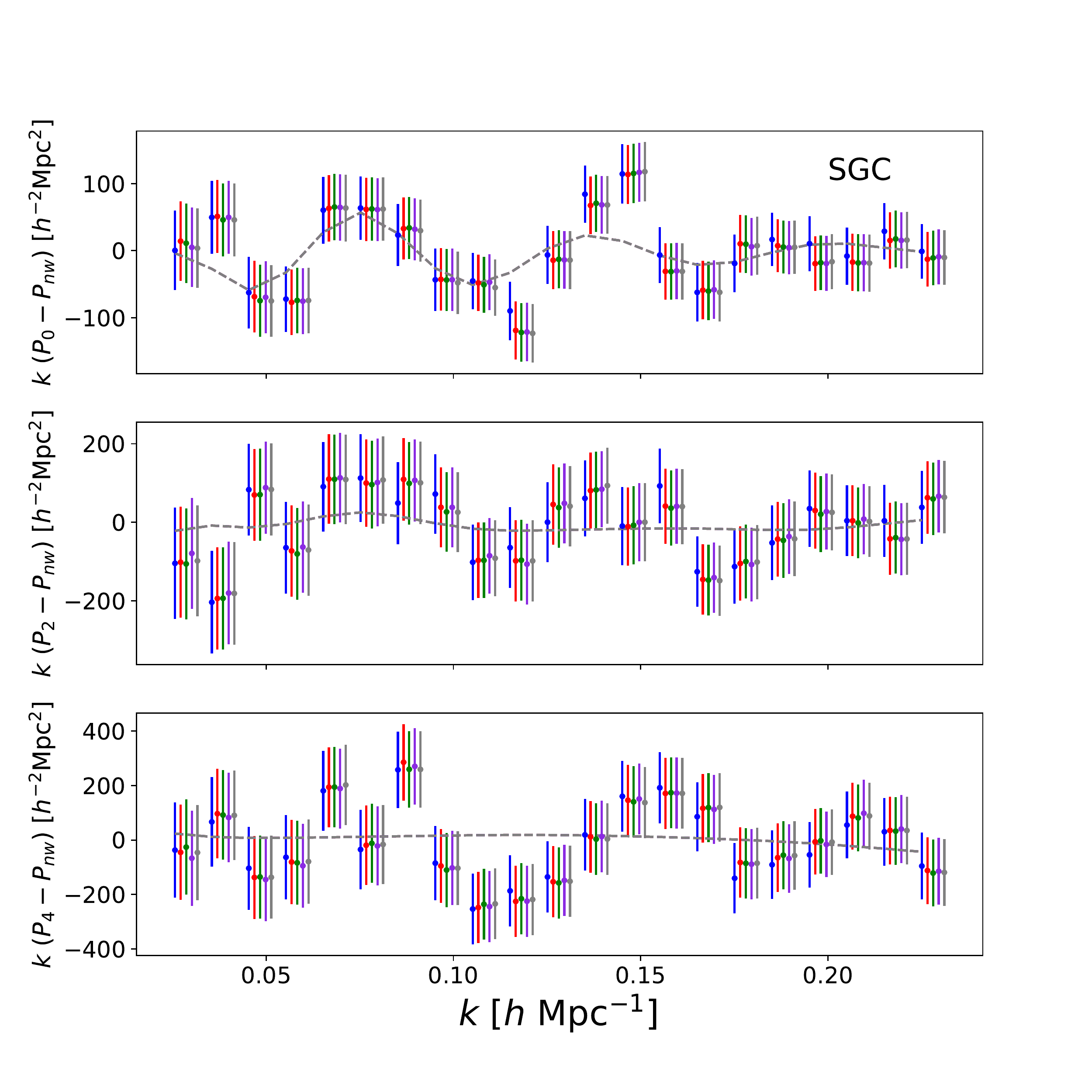}
    \caption{The effect of the NN weights in the power spectrum in comparison to the default weight (blue), as a function of various setups of the NN method (Table \ref{tab:alphatable}). We show the isolated BAO wiggles in the power spectra of the H256 catalogues. The top set of panels shows the BAO signal in the NGC and the bottom set shows the SGC. Points are the data with error bars derived from the covariance matrix and the solid lines are the best-fit models. The wiggles are isolated by subtracting the power spectra by a smoothed spectra, which is produced from fitting the default catalogue to a smoothed model.  Data points are shifted horizontally to prevent overlap, but best-fits lines are not shifted. Colors correspond to the catalogues with different systematic weights: red, gree, purple and grey representing known-1z, known-2z, all-1z, and all 2-z, respectively.}
    \label{fig:BAOwig256}
\end{figure}

We include here Figure \ref{fig:BAOwig256}, illustrating the effect of different NN weights in the power spectrum of the QSO catalogue.  The colors correspond to the same weights as in Figure \ref{fig:BAOwig512}.  We see a general trend of points not being shifted more than 0.5$\sigma$, with one point shifted by a full $\sigma$.

\section{High-z Sample}
\label{App:highz}
The QSO catalogues released in SDSS-IV DR16 includes the sample over $0.8<z<3.5$. The main QSO sample that \citet{Neveux20} and this paper analyzed covers objects in the redshift range $0.8<z<2.2$, with an effective redshift of $z=1.48$. However, the deeper portions of the sample have not been examined closely to date.
There is a potential to extract cosmological information from these high-z QSOs despite the poor signal-to-noise ratio of this sample.  Here we present the BAO analysis of the high-z sample using the NN weights for this sample from \citet{rezaie2021}.

A particular challenge with this sample is the lack of mocks to supplement the data, as the EZmocks we used in the analysis of the standard sample do not cover the higher redshifts of this high-z sample.  Therefore we rely on a simple Gaussian approximations for the covariance matrix.  Our goal is to asses the impact of including the high-z sample in extracting the BAO signal.  While in principal we could extract a BAO signal from the high-z sample alone, the signal-to-noise proves to be too poor to give a meaningful BAO result.  In particular, the quadrupole of the power spectrum is extremely noisy.  

\begin{figure}
    \centering
    \includegraphics[width=\columnwidth]{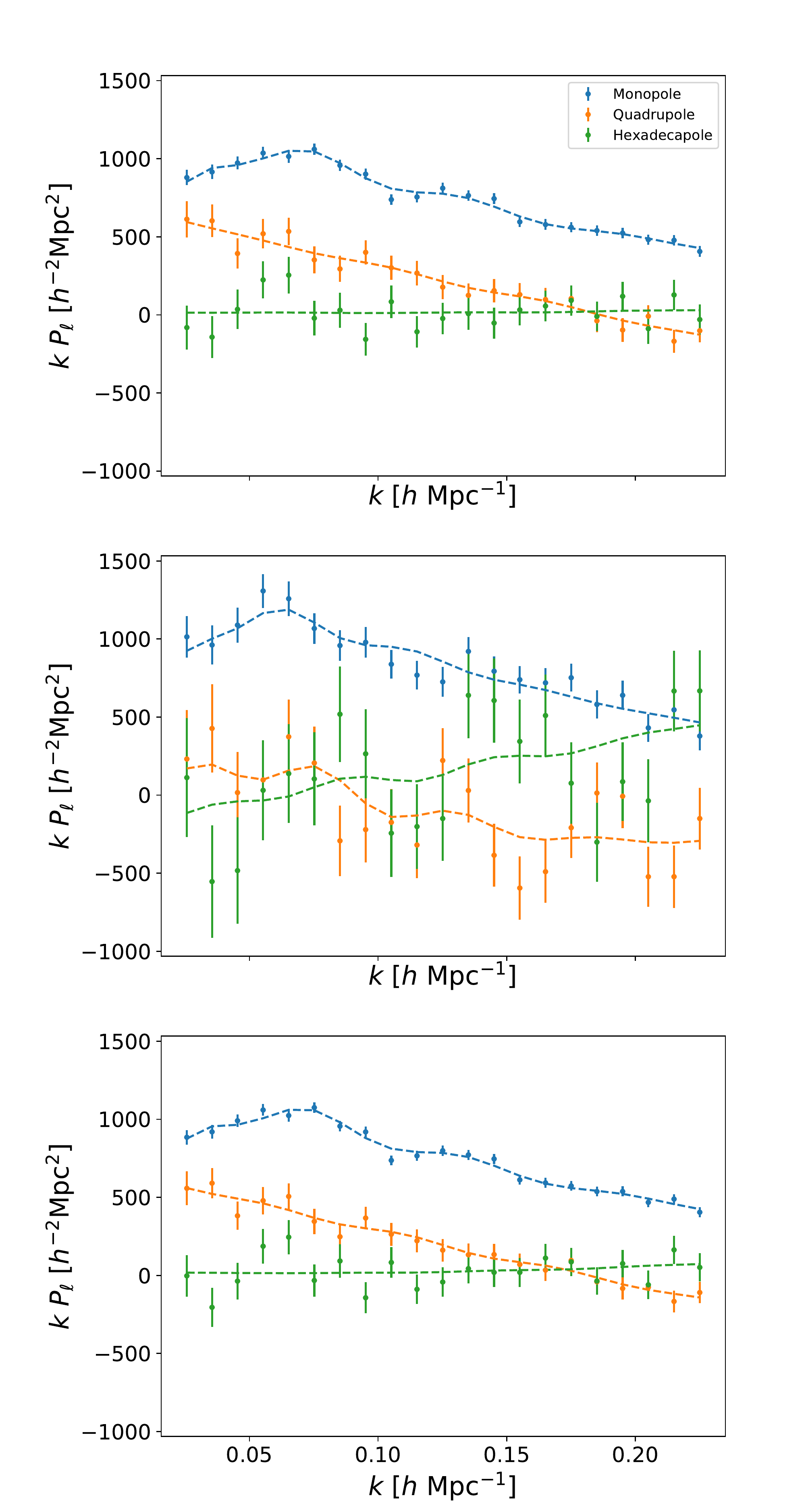}
    \caption{The power spectra of the NGC calculated using different samples: standard 0.8<z<2.2 (top), highz 2.2<z<3.5 (middle) and combined 0.8<z<3.5 (bottom).  The error is derived from the default covariance matrix, and the highz and combined samples also utilize a Gaussian approximation. The dashed lines indicate the best-fit model.  The highz sample has a much lower signal-to-noise ratio.}
    \label{fig:NGC_samples}
\end{figure}

To construct the covariance matrix for the high-z sample, we calculate a Gaussian covariance for the standard sample and compare the BAO fit using this covariance with the fit using the covariance derived from the mocks.  After finding a reasonable agreement between the two, we rescale the Gaussian covariance matrix of the main sample with the ratio of the shot noises between the two samples, as the shot noise is the main source of statistical error in the QSO sample. Figure \ref{fig:NGC_samples} shows the result of fitting the NGC of the different redshift samples: standard ($0.8<z<2.2$) with the mock covariance matrix, high-z ($2.2<z<3.5$) with the Gaussian covariance matrix, and when combined ($0.8<z<3.5$). 

\begin{table}
\centering
\caption[h]{Results of fitting with the highz sample.  We examine the case of catalogs treated with the known-1z-H512 mitigation and calculate spectra for the standard sample and the standard+highz sample for each galactic cap.}
\resizebox{1.0\columnwidth}{!}{% use resizebox with textwidth
\begin{tabular}{ l |  l  l  l }
 & $\alpha_\parallel$ & $\alpha_\perp$ & $\chi^2$ / dof\\
  \hline
   NGC & & & \\
   main & 1.016 $\pm$ 0.0788 &  1.009 $\pm$ 0.0367  & 41.14 / 50 \\
   combined & 0.9976 $\pm$ 0.0712 & 1.019 $\pm$ 0.0370 & 46.13 / 50 \\
   \hline
   SGC & & & \\
   main & 1.054 $\pm$ 0.0606 &  1.017 $\pm$ 0.0909  & 45.24 / 50 \\
   combined & 1.040 $\pm$ 0.0539 & 1.000 $\pm$ 0.0719 & 46.10 / 50 \\
   \hline
   NGC+SGC & & & \\
   main & 1.030 $\pm$ 0.0552 &  1.014 $\pm$ 0.0336  & 90.80 / 102 \\   
   combined & 1.012 $\pm$ 0.050 &  1.018 $\pm$ 0.0299  & 94.91 / 102 \\  
   \hline

\end{tabular}% close resizebox
}
\label{tab:highztable}
\end{table}

In Table \ref{tab:highztable} we show the AP parameters obtained from the fits.  The growth parameter $f$ is freed for all fits.  We do not include the window function in the fits of the combined sample, but note that the window function was found to shift the best-fit $\alpha$s by 0.1-0.2\% in \citet{Neveux20}. As the table indicates, the high-z sample on its own does not constrain the BAO signal. With the combined sample, we find around a 10\% reduction in the error on $\alpha_\parallel$ and $\alpha_\perp$. 

%%%%%%%%%%%%%%%%%%%%%%%%%%%%%%%%%%%%%%%%%%%%%%%%%%

% Don't change these lines
\bsp	% typesetting comment
\label{lastpage}
\end{document}